\documentclass[manuscript]{aastex61}

\usepackage{amsmath}
\usepackage{bm}

\def\vec#1{\ensuremath{\bm{#1}}}
 \newcommand{\mathd}{\ensuremath{{\rm d}}}
\received{}
\revised{}
\accepted{}

\shorttitle{Kink oscillations in coronal loops with elliptic cross-sections}
\shortauthors{Guo et al.}

\begin{document}

\title{Kink Oscillations in Solar Coronal Loops with Elliptical Cross-Sections. I. the linear regime}

\correspondingauthor{Mingzhe Guo}
\email{m.guo@sdu.edu.cn}

\author{Mingzhe Guo}
\affiliation{Shandong Provincial Key Laboratory of Optical Astronomy and Solar-Terrestrial Environment, Institute of Space Sciences, Shandong University, Weihai 264209, China}

\author{Bo Li}
\affiliation{Shandong Provincial Key Laboratory of Optical Astronomy and Solar-Terrestrial Environment, Institute of Space Sciences, Shandong University, Weihai 264209, China}

\author{Tom Van Doorsselaere}
\affiliation{Centre for mathematical Plasma Astrophysics, Department of Mathematics, KU Leuven, 3001 Leuven, Belgium}




\begin{abstract}
The cross sections of solar coronal loops are suggested to be rarely circular. We examine linear kink oscillations in straight, density-enhanced, magnetic cylinders with elliptical cross-sections by solving the three-dimensional magnetohydrodynamic equations from an initial-value-problem perspective. Motivated by relevant eigen-mode analyses, we distinguish between two independent polarizations, one along the major axis (the M-modes) and the other along the minor one (the m-modes). We find that, as happens for coronal loops with circular cross-sections, the apparent damping of the transverse displacement of the loop axis is accompanied by the accumulation of transverse Alfv\'enic motions and the consequent development of small-scales therein, suggesting the robustness of the concepts of resonant absorption and phase-mixing. In addition, two stages can in general be told apart in the temporal evolution of the loop displacement; a Gaussian time dependence precedes an exponential one. For the two examined density ratios between loops and their surroundings, the periods of the M-modes (m-modes) tend to increase (decrease) with the major-to-minor-half-axis ratio, and the damping times in the exponential stage 
for the M-modes tend to exceed their m-mode counterparts. This is true for the two transverse profiles we examine. However, the relative magnitudes of the damping times in the exponential stage for different polarizations depend on the specification of the transverse profile and/or the density contrast. The applications of our numerical findings are discussed in the context of coronal seismology. 

\end{abstract}


\section{INTRODUCTION} 
\label{sec_intro}

Cyclic transverse displacements of solar coronal loops have been amply observed 
   and have been customarily interpreted as 
   kink waves collectively supported therein 
   \citep[see e.g.,][for recent reviews]{2005LRSP....2....3N,2007SoPh..246....3B,2012RSPTA.370.3193D,2016GMS...216..395W,2016SSRv..200...75N}.
While evidence for propagating kink waves has been offered by 
   instruments like the Coronal Multi-Channel Polarimeter
   (CoMP; 
   e.g., \citeauthor{2007Sci...317.1192T}~\citeyear{2007Sci...317.1192T},
   \citeauthor{2009ApJ...697.1384T}~\citeyear{2009ApJ...697.1384T}),
   most of the transverse motions have been found to be compatible 
   with standing kink waves (or kink oscillations in other words)
   ever since they were first imaged by the Transition Region and Coronal Explorer
       (TRACE,
       \citeauthor{1999ApJ...520..880A}~\citeyear{1999ApJ...520..880A}
       and
       \citeauthor{1999Sci...285..862N}~\citeyear{1999Sci...285..862N})
       and subsequently by Hinode \citep{2008A&A...487L..17V, 2008A&A...482L...9O, 2008A&A...489L..49E},
       the Solar TErrestrial RElations Observatories (STEREO,
       \citeauthor{2009ApJ...698..397V}~\citeyear{2009ApJ...698..397V}
       )
       and the Solar Dynamics Observatory/Atmospheric Imaging Assembly 
       (SDO/AIA, e.g., 
       \citeauthor{2011ApJ...736..102A}~\citeyear{2011ApJ...736..102A};
       \citeauthor{2012A&A...537A..49W}~\citeyear{2012A&A...537A..49W}).  
Two regimes exist as far as kink oscillations are concerned.
The loop displacements associated with the so-called decayless kink oscillations 
    barely exceed the loop diameter and
    experience little damping~\citep[e.g.,][]{2012ApJ...751L..27W, 2012ApJ...759..144T, 2013A&A...560A.107A, 2013A&A...552A..57N, 2019ApJ...884L..40A}.
In addition, they tend not to be connected with eruptive events like flares 
    or coronal mass ejections~\citep[e.g.,][]{2015A&A...583A.136A}. 
This then raises
    the question as to whether the continuous energy supply, necessary for maintaining
    a nearly constant oscillatory behavior and most likely connected with footpoint motions, 
    comes in a quasi-steady~\citep[e.g.,][]{2016A&A...591L...5N},
    nearly monochromatic 
        (e.g., \citeauthor{2017A&A...604A.130K}~\citeyear{2017A&A...604A.130K},
               \citeauthor{2019ApJ...870...55G}~\citeyear{2019ApJ...870...55G}; 
         see also \citeauthor{2016ApJ...830L..22A}~\citeyear{2016ApJ...830L..22A}),
    or random manner~\citep{2020A&A...633L...8A}.
Decaying kink oscillations, on the other hand, are of larger amplitude and
    almost always associated with
    lower coronal eruptions~\citep[e.g.,][]{2015A&A...577A...4Z,2019ApJS..241...31N}.
They tend to damp in several cycles, a feature evident in both individual
    measurements~\citep[e.g.,][]{1999ApJ...520..880A, 1999Sci...285..862N}
    and statistical surveys~\citep[e.g.,][]{2013A&A...552A.138V, 2016A&A...585A.137G,2019ApJS..241...31N}.

Decaying kink oscillations, the focus of the present study,
    have been routinely exploited from the perspective of coronal seismology
    since their discovery
    (see e.g., the early review by~\citeauthor{2000SoPh..193..139R}~\citeyear{2000SoPh..193..139R}).
This is understandable because the measured periods can be inverted for 
    the magnetic field strength in coronal loops, which is difficult
    to routinely measure otherwise
    (\citeauthor{2001A&A...372L..53N}~\citeyear{2001A&A...372L..53N},
    also the reviews by e.g., 
    \citeauthor{2005LRSP....2....3N}~\citeyear{2005LRSP....2....3N},
    and \citeauthor{2016SSRv..200...75N}~\citeyear{2016SSRv..200...75N}). 
The measured damping times have proven equally useful given that the damping is largely accepted to
    result from the resonant conversion of the kink energy to localized Alfv\'en waves,
    a concept originally proposed to account for coronal heating
    \citep[e.g.,][]{1978ApJ...226..650I,1979ApJ...227..319W,1988JGR....93.5423H}
    and later invoked for seismological purposes
    (\citeauthor{2002ApJ...577..475R}~\citeyear{2002ApJ...577..475R},
     \citeauthor{2002A&A...394L..39G}~\citeyear{2002A&A...394L..39G},
     \citeauthor{2003ApJ...598.1375A}~\citeyear{2003ApJ...598.1375A},
     \citeauthor{2004ApJ...606.1223V}~\citeyear{2004ApJ...606.1223V};
     see also the review by 
     \citeauthor{2011SSRv..158..289G}~\citeyear{2011SSRv..158..289G}
     and references therein).
When put into practice, the theory of this ``resonant absorption" enables
     one to constrain the transverse inhomogeneity lengthscale of the loop density
     either analytically~\citep[e.g.,][]{2008A&A...484..851G}
     or largely numerically~\citep[e.g.,][]{2007A&A...463..333A, 2014ApJ...781..111S}.
Note that the relevant theoretical analysis is usually conducted from
     the eigen-value problem perspective~\citep[see e.g., the review by][]{2011SSRv..158..289G}.
It then follows from the pertinent solution to the initial-value problem
     that the damping rate pertains to the asymptotic stage where the system
     has evolved for a time much longer than $P_{\rm k}/2\pi$ 
     and the wave damping is exponential in time,
     where $P_{\rm k}$ is the kink period
     \citep[e.g.,][]{2002ApJ...577..475R,2006JPlPh..72..285R}.
Fortunate is that an exponential damping profile is usually present in observations of loop oscillations, 
     but unfortunate is that the damping times thus measured, together with the measured periods,
     allow one to constrain the equilibrium loop parameters only to a one-dimensional (1D)
     curve in the 3D parameter space formed by the longitudinal Alfv\'en time, the density contrast between the loop and its ambient surroundings, and the transverse density lengthscale
     (e.g., 
     \citeauthor{2007A&A...463..333A}~\citeyear{2007A&A...463..333A},
     \citeauthor{2008A&A...484..851G}~\citeyear{2008A&A...484..851G},
     \citeauthor{2015ApJ...812...22C}~\citeyear{2015ApJ...812...22C};
     see also the comments on the under-determined nature of the inversion problem
     by \citeauthor{2019A&A...622A..44A}~\citeyear{2019A&A...622A..44A}).

The under-determined situation in seismological practices 
    has been shown to improve when one incorporates the information
    in the transitory phase, namely the phase before the system evolves
    into the asymptotic stage. 
This was first noted in the numerical studies on propagating kink waves by
    \citet{2012A&A...539A..37P}, where the authors found that
    resonant absorption operates in this case as well but 
    in general the spatial damping profile is Gaussian-like in the first couple of wavelengths
    before becoming an exponential one. 
The numerical findings soon received the theoretical support from~\citet{2013A&A...551A..39H}.
In addition, a temporal damping profile, in which a Gaussian time dependence
    precedes an exponential one, is expected for kink oscillations if one translates 
    the spatial behavior of propagating waves to the temporal behavior of standing waves
    with the aid of the axial group speed~\citep{2013A&A...551A..40P}. 
Theoretically speaking, 
    this expectation was corroborated by analytical~\citep{2013A&A...555A..27R} 
    and numerical studies~\citep[e.g.,][]{2015ApJ...803...43S, 2016A&A...595A..81M}.
More importantly, observational evidence for the Gaussian damping 
    was indeed found for decaying kink oscillations~\citep{2016A&A...585L...6P}.
As such the damping times in the Gaussian stage become an additional measurable
    that further constrains the loop parameters~\citep[e.g.,][]{2016A&A...589A.136P,2018ApJ...860...31P}.   
Recently the damping profiles including the Gaussian stage in coronal loops with various density profiles were provided in a look-up table, which can be used for seismology~\citep{2019FrASS...6...22P}.

The theories behind all the afore-mentioned sesmological applications have 
    assumed that coronal loops are monolithic density-enhancements
    that are embedded in an otherwise uniform ambient 
    and straight magnetic fluxtubes with a constant circular cross-section everywhere~\citep[e.g.,][]{2011SSRv..158..289G}. 
This latter geometrical assumption actually involves
    two interconnected aspects: the loop shape and the loop cross-sectional properties. 
The loop shape, namely the morphology of the loop
    axis projected onto the plane of the sky (PoS), 
    is certainly curved rather than being
    straight~\citep[see e.g., the review by][and references therein]{2009SSRv..149...31A}. 
Considerably less certain are the cross-sectional properties, by which we mean the spatial
    variation along the loop of the surface enclosing the density enhancement.
While stereoscopic observations with, say, the twin STEREO spacecraft can in principle
    place constraints on this aspect~\citep[e.g., the review by][]{2011LRSP....8....5A},
    a practical implementation is not straightforward, the primary issue being
    the optical thinness of the corona in the relevant passbands
    \citep[see e.g.,][]{2013ApJ...775..120M}.
As for spectroscopic measurements, issues associated with determining the filling factors
    and the loop background further complicate the problem~\citep[e.g.][and references therein]{2019ApJ...885....7K}. 
A common practice is then to employ some modeled coronal magnetic field
    to examine the geometry of magnetic fluxtubes.
Note that, by construction, the cross-section is allowed to be prescribed at only 
    one location but needs to be computed elsewhere~\citep[e.g.][]{2009A&A...506..885R}.      
While it is statistically true that cross-sectional areas tend to vary more strongly
    in current-free fields than in force-free but non-current-free fields
    \citep[e.g.,][and references therein]{1998PASJ...50..111W, 2000ApJ...542..504K,
    	2006ApJ...639..459L},
    there seems to be no definitive conclusion as to the cross-sectional behavior of individual fluxtubes.         
Depending on the choice of the field model, even if 
    the cross-sectional area of selected fluxtubes
    may not vary too much~\citep[e.g.,][]{2013ApJ...775..120M}, the dimensions along different directions intersecting the cross-sections may do so~\citep[e.g.,][]{1998PASJ...50..111W}.
It is even possible, at least for thin fluxtubes in planar current-free fields,
    that both the shape and dimensions of
    fluxtube cross-sections are preserved~\citep{2015SoPh..290..423R}.

Now return to kink oscillations in coronal loops. 
With the realistic features of magnetic fluxtubes in mind, 
    it should be ideal for one to set up an equilibrium
    that addresses simultaneously a curved loop shape and
    a position-dependent non-circular cross-section.
A number of numerical studies have indeed been devoted to this purpose
    \citep[e.g.,][]{2008ApJ...682.1338M, 2010ApJ...714..170S, 
    	2011ApJ...728...87S,2020ApJ...894L..23M}.
However, one issue associated with this practice is that
    the wave behavior is at best marginally
    analytically tractable~\citep{2009A&A...506..885R}, thereby
    making it difficult to draw a definitive conclusion on such issues 
    as the relative importance of wave leakage
    versus resonant absorption for damping kink oscillations.
It is then no surprise to see that more studies tend to isolate one geometrical factor
    out of many.
For instance, a curved loop with a position-independent circular cross-section 
    has been rather extensively examined
    (see the review by~\citeauthor{2009SSRv..149..299V}~\citeyear{2009SSRv..149..299V}
    for earlier papers, and also 
    \citeauthor{2009ApJ...699L..72D}~\citeyear{2009ApJ...699L..72D},
    \citeauthor{2014ApJ...784..101P}~\citeyear{2014ApJ...784..101P}
    ).
Likewise, a straight loop with a position-independent elliptic cross-section
    has also received much attention given that it permits a largely analytical
    eigen-mode analysis 
    (e.g., \citeauthor{2003A&A...409..287R}~\citeyear{2003A&A...409..287R},
           \citeauthor{2009A&A...494..295E}~\citeyear{2009A&A...494..295E},
           \citeauthor{2011A&A...527A..53M}~\citeyear{2011A&A...527A..53M};
           also Verth et al., private communications).
It was shown that this geometry, although different from the classical
    one only in replacing a circular
    with an elliptic cross-section, brings forth the important difference that
    now kink oscillations polarized along the major and minor axes 
    are no longer degenerate.
Nonetheless, resonant absorption was found to operate for kink oscillations
    with these two independent polarizations~\citep[][hereafter R03]{2003A&A...409..287R}.

This manuscript is intended to examine kink oscillations in active region (AR) loops
    modeled as straight, density-enhanced, fluxtubes with elliptic cross-sections
    by numerically solving 
    the three-dimensional (3D) ideal MHD equations.
Some justifications on our approach seem necessary
    at this point.
First, while our equilibrium is similar to the one in the eigen-mode analysis by R03, 
    examining kink oscillations from the initial-value-problem perspective 
    will allow us to address a number of additional questions.
For instance, can a Gaussian damping envelope be identified before an exponential one
    as happens for loops with circular cross-sections? 
What will be the seismological applications if both phases occur in general,  
    and will it be possible to infer the aspect ratio of the cross-sections in particular?
Second, neglecting the curvature of the loop axis
    means that the wave damping will be solely due to
    resonant absorption taking place primarily in the neighborhood of the loop boundary.
This can be only partially justified because while there are studies suggesting
    that resonant absorption dominates such curvature-related factors as
    wave leakage or additional resonant absorption taking place in the ambient corona
    \citep{2006ApJ...650L..91T},
    there are also studies that indicate otherwise
    \citep[e.g.,][]{2008ApJ...682.1338M,2011ApJ...726...42S}.     
Having said that, loop curvature is expected to influence 
    the periods to a lesser degree, at least when
    a position-independent circular cross-section is adopted for the modeled loops
    (\citeauthor{2006ApJ...650L..91T}~\citeyear{2006ApJ...650L..91T}; 
     see also the review by 
     \citeauthor{2009SSRv..149..299V}~\citeyear{2009SSRv..149..299V}).
Now thinking about loops for which the loop axis is curved but 
    a position-independent elliptic cross-section can be maintained, 
    our results regarding the periods are likely to apply, even though
    the computed damping rates need to be treated with some caution. 
  
This manuscript is organized as follows. 
Section~\ref{sec_NumModel} details 
    the specification of our equilibrium configuration,
    and our numerical implementation.
The numerical results are then described in Section~\ref{sec_num_results},
    followed by some rather detailed discussions on their potential seismological
    applications in Section~\ref{sec_discus}.
Section~\ref{sec_summary} summarizes the present study, ending 
    with some concluding remarks.

\section{Model Formulation and Numerical Methods}
\label{sec_NumModel}

\subsection{Equilibrium Setup}
\label{sec_sub_IC}
We work in the framework of ideal MHD throughout, for which the primary dependents 
   are the mass density $\rho$, velocity $\vec{v}$, magnetic field $\vec{B}$, and
   thermal pressure $p$.
Assuming an electron-proton plasma, 
   one then relates the mass density to the electron number density $N$
   through $\rho = N m_p$ with $m_p$ being the proton mass.
Likewise, the thermal pressure is related to the electron temperature $T$ 
   via $p = 2 N k_B T$ where $k_B$ is the Boltzmann constant.     
   
We start with constructing an equilibrium for which all physical parameters are denoted by 
   a subscript $0$.
A Cartesian coordinate system $(x,y,z)$ is adopted throughout, and 
   the equilibrium magnetic field $\vec{B}_0$ is assumed
   to be everywhere in the $z$-direction.  
We model a coronal loop as a static, field-aligned, density enhancement, for which the axis
   is aligned with the $z$-axis and the ambient corona is homogeneous. 
This loop is of length $L$, as follows from the assumption that the loop is bounded
   at both $z=0$ and $z=L$ by two photospheres. 
The equilibrium parameters are assumed to be structured only in the transverse directions.   
Let the subscripts ${\rm i}$ and ${\rm e}$ denote the equilibrium  
   parameters at the loop axis and far from the loop, respectively.   
We then assume that $\rho_0$ takes the form
\begin{eqnarray}
\rho_0(x, y) = \rho_{\rm e}+(\rho_{\mathrm i}-\rho_{\mathrm e})f(x, y)~~,
\label{eq_rhoEq}
\end{eqnarray}
   which describes a density profile varying continuously from
   the internal value at the axis to the external value
   in the sufficiently far ambient medium. 
This variation is realized through the function $f$, which depends on
   the transverse coordinates $x$ and $y$ through an
   intermediate variable $\bar{r}$ as
\begin{eqnarray}
\displaystyle
f(x, y) \equiv f(\bar{r}) = \frac{1}{1+\bar{r}^\alpha}~~,
\label{eq_fr}
\end{eqnarray}
   with 
\begin{equation}
\bar{r} = \sqrt{\left(\frac{x}{a}\right)^2+\left(\frac{y}{b}\right)^2}~.
\label{eq_r}
\end{equation}
We further assume that $a \ge b$.
As such, $f(x,y)$ yields an ecliptic cross-section with a mean major (minor) half-axis 
   being $a$ ($b$).
Furthermore, the parameter $\alpha$ characterizes how steep this $f$ varies. 
Instead of specifying the distribution of the pressure, we assume that the
   equilibrium electron temperature ($T_0$)
   follows the same spatial dependence as the mass density.
The thermal pressure $p_0(x, y)$ is then readily derived, and
   the magnetic field strength $B_0$ follows from the transverse force balance,
\begin{eqnarray}
\displaystyle
  p_0(x, y)   + \frac{B_0^2(x, y)}{2\mu_0}
= p_{\rm i}   + \frac{B^2_{\rm i}}{2\mu_0}~~,
\label{eq_forcebalance}
\end{eqnarray}
   where $\mu_0$ is the magnetic permeability in free space.

\begin{figure}
	\centering
	\gridline{\fig{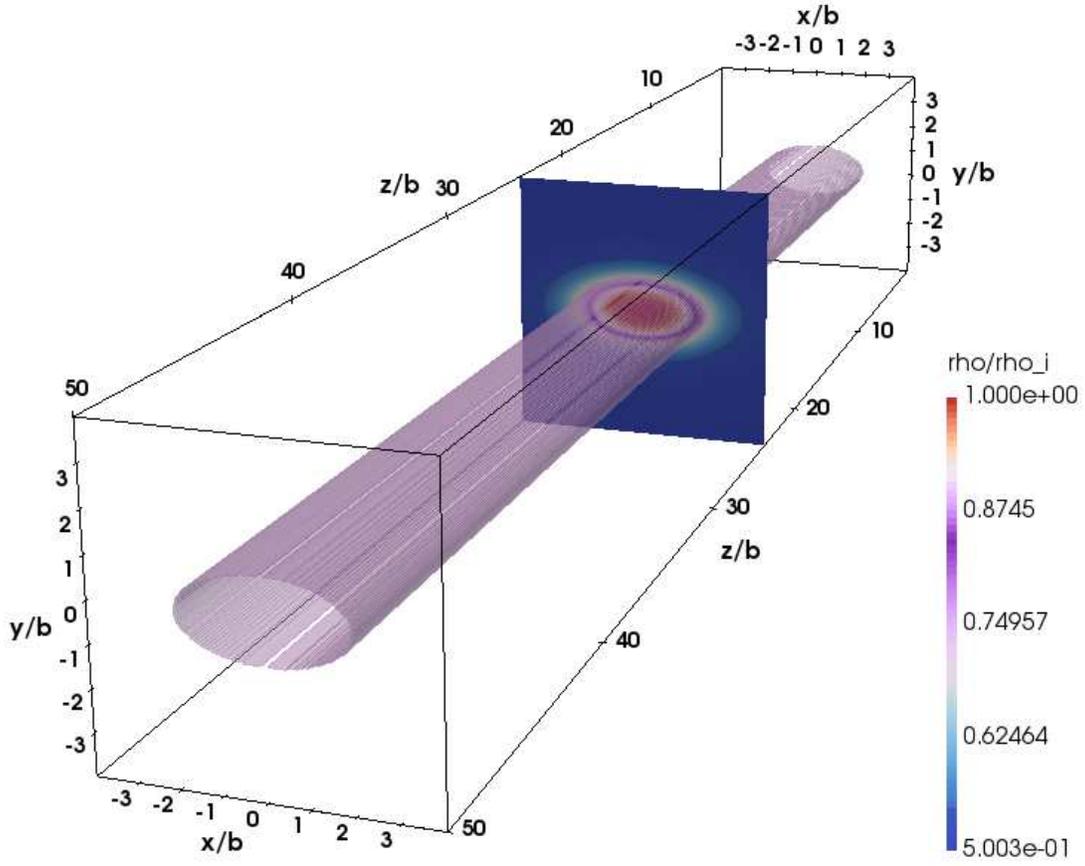}{0.8\textwidth}{}}
	\caption{
	Illustration of the equilibrium configuration involving a straight coronal loop
	    with elliptic cross-sections realized through Equation~\eqref{eq_fr}.
	The loop boundary with a mean major half-axis $a$ and minor half-axis $b$
	    is given by the shaded surface.
	The filled contours are for the distribution of the mass density
	    in the transverse plane.  
	For illustration purposes, here we adopt an aspect ratio $a/b=2$,
	    an internal-to-external density ratio $\rho_{\rm i}/\rho_{\rm e}=2$,
	    a steepness parameter $\alpha = 5$, 
	    and a length-to-minor-half-axis ratio $L/b = 50$.     
	}			
	\label{fig_equil}
\end{figure}

The equilibrium configuration is fully determined once we specify
   the geometrical parameters ($L$, $a$, $b$, and $\alpha$)
   and the physical parameters 
   ($\rho_{\rm i}$, $\rho_{\rm e}$, $T_{\rm i}$, $T_{\rm e}$, 
   and $B_i$).
Alternatively, these parameters can also be grouped into
   a set of dimensional parameters $[b; \rho_{\rm i}, T_{\rm i}; B_{\rm i}]$ 
   and a set of dimensionless ones 
   $[L/b, a/b, \alpha; \rho_{\rm i}/\rho_{\rm e}, T_{\rm i}/T_{\rm e}]$. 
Evidently, there are too many parameters to explore with 3D simulations,
   and therefore we choose to fix many.    
To be specific, we fix the minor half-axis ($b$) at $1000$~km,
   the length-to-minor-half-axis ratio ($L/b$) at $50$,
   and the steepness parameter $\alpha$ at $5$.
The electron temperature at the loop axis $T_{\rm i}$ is taken
   to be $1.4$~MK, and the internal-to-external temperature ratio 
   $T_{\rm i}/T_{\rm e}$ is fixed at $2$.
In addition, we fix the internal density $\rho_{\rm i}$ such that it corresponds to
   an electron number density of $10^{9}$~cm$^{-3}$.
The magnetic field strength at the loop axis ($B_{\rm i}$) is taken to be $15$~G, 
   resulting in a plasma $\beta$ of $\beta_{\rm i} = 0.043$
   and an Alfv\'en speed of $v_{\rm Ai} = 1035$~km~s$^{-1}$.   
We are left with the aspect ratio ($a/b$) and
   the density ratio $\rho_{\rm i}/\rho_{\rm e}$ to vary.   
The parameters we either fix or allow to vary are largely typical of
   warm active region loops~\citep[e.g., the review by][]{2014LRSP...11....4R},
   even though    
   the steepness parameter is known to be difficult 
   to observationally constrain~\citep[e.g.,][]{2003ApJ...598.1375A,2018ApJ...860...31P}.
For illustration purposes, Figure~\ref{fig_equil} shows
   an equilibrium loop corresponding to the combination of 
   $[a/b, \rho_{\rm i}/\rho_{\rm e}]  = [2, 2]$. 
The loop boundary is shown by the shaded surface, and
   the filled contours at the loop apex represent the transverse distribution
   of the mass density.

\subsection{Numerical Setup}
\label{sec_sub_numsetup}

We examine the properties of the fundamental kink modes supported by the considered equilibrium
   by numerically following the response of the equilibrium to an initial
   velocity perturbation $\vec{v}(x, y, z; t=0)$. 
However, following R03, we know a priori that now the kink modes
   correspond to two independent polarizations as far as the velocity field is concerned.
This means that two different types of initial perturbations 
   need to be distinguished from the outset.
Accordingly, different setups for the numerical grid are necessary
   to save computational costs.      
For the ease of description, let the modes polarized
   in the direction of the major (minor) axis be denoted by
   the M-modes (m-modes). 
In addition, recall that the major and minor axes are aligned with
   the $x$- and $y$- directions, respectively.   
      
To excite fundamental M-modes, 
   we introduce an initial perturbation of the form 
\begin{eqnarray}
\displaystyle
   \vec{v}({x, y, z; t=0}) 
  = v_{0} f(x, y) \sin\left(\frac{\pi z}{L}\right)\hat{x}~~.
\label{eq_vx}
\end{eqnarray}
Here $v_0$ denotes the amplitude of the perturbation,
     and $\hat{x}$ represents the unit vector in the $x$-direction.
In addition, $f(x, y)$ has been described by Equation~\eqref{eq_fr},
     and is adopted here to avoid introducing additional parameters
     that characterize the spatial profile of initial perturbations. 
The symmetric properties of the kink modes in this situation allow 
     us to consider only a quarter of 
     the nominal computational domain of
     $(-\infty, \infty) \times (-\infty, \infty) \times [0, L]$.
In practice, we adopt a simulation domain of
     $[-15, 15]~b \times [0,   15]~b \times [0, L/2]$, 
     and apply symmetric boundary conditions 
     at both $y=0$ and
      $z=L/2$~\citep[e.g.,][]{2015ApJ...809...72A, 2017A&A...604A.130K}. 
Our base computations employ $25$ uniform cells in the $z$-direction,
     but adopt $590$ ($175$) non-uniformly spaced cells
     in the $x$- ($y$-) direction.
{ To be specific, along the $x$-direction, 
     480 grid points are uniformly distributed
     for $-8b\leq x\leq 8b$, beyond which a stretched grid is used. 
Likewise, 120 uniform grids are adopted 
     in the $y$-direction in the interval $0\leq y\leq 4b$,
     outside which a stretched grid is employed.
This grid system is constructed primarily to make our computations less numerically
     expensive, with the uniform portion adopted to adequately capture
     the wave dynamics.}
The resulting spatial resolution reaches up to $\sim b/30 = 33~$km
     in both directions.   
     
To excite fundamental m-modes, we introduce an initial perturbation identical in form
     to Equation~\eqref{eq_vx} except that $\hat{x}$ is replaced with $\hat{y}$.
We adopt a simulation domain of      
     $[0,   15]~b \times [-15, 15]~b \times  [0, L/2]$, 
     and apply symmetric boundary conditions 
     at both $x=0$ and $z=L/2$. 
In our base computations, 
     the grid setup in the $z$-direction is identical to the M-mode case.
However, now $295$ and $350$ non-uniformly distributed cells
     are used in the $x$- and $y$- directions, 
     respectively. 
 {We employ a uniform grid of 240 (240) points in the interval of 
     $0\leq x\leq 8b$ ($-4b\leq y\leq 4b$) but use a stretched grid in the rest of the domain.}
This choice ensures the same highest resolution as in the base computations for
     the M-modes.

The rest of the numerical implementation is common to both polarizations. 
The amplitude of the initial perturbation $v_0$ is taken to be $4$~km~s$^{-1}$
     (or equivalently $\sim 0.0039~v_{\rm Ai}$) to avoid complications
     associated with non-linearities. 
Regarding the boundary conditions at the loop footpoint ($z=0$), 
     both $v_x$ and $v_y$ are set to vanish, 
     while $v_z$, $B_x$, and $B_y$ follow the zero-gradient condition.
The remaining physical variables are fixed at their initial values here.
Outflow boundary conditions are applied to all lateral boundaries.  
We then solve the time-dependent, ideal, MHD equations
     with the PLUTO code \citep{2007ApJS..170..228M}.
Following the reconstruct-solve-average strategy, 
     we choose the piecewise parabolic scheme for reconstruction,
     the HLLD Riemann solver for computing the numerical fluxes,
     and the second-order Runge-Kutta method for time marching. 
 
\section{Numerical Results}
\label{sec_num_results}
We examine a number of combinations of $a/b$ and $\rho_{\rm i}/\rho_{\rm e}$.
The aspect ratio ($a/b$) varies between $1$ and $3$, a range compatible with
   the spectroscopically derived values~\citep{2019ApJ...885....7K}.
Alternatively, the deviation of a cross-section from a circular one can also be measured in terms of flattening $\xi = 1-b/a$. 
For an $a/b=3$,
  the flattening reads $\xi=2/3$.
Given the computational costs of 3D simulations, we examine only two values of 
   the density ratio ($\rho_{\rm i}/\rho_{\rm e}$), one being $2$ and 
   the other being $8$.
On the one hand, these two density ratios lie almost at the extreme of
   the observed range for typical
   active region loops
   ($2-10$, e.g., \citeauthor{2004ApJ...600..458A}~\citeyear{2004ApJ...600..458A}),
   and therefore are likely to be representative of low- and high-density ratios.  
On the other hand, as can be deduced from R03, the damping times for the m-modes
   tend to exceed those for the M-modes when $\rho_{\rm i}/\rho_{\rm e} \lesssim 5$, 
   whereas this trend is reversed when $\rho_{\rm i}/\rho_{\rm e} \gtrsim 5$. 
Note that our prescription of the transverse distribution (Equation~\ref{eq_fr}) is
   different from R03, where a linear profile is implemented with the
   elliptic coordinates (Equation~\ref{eq_app_fxy}).
In addition, R03 worked in the framework
   of zero-$\beta$ MHD, and adopted the thin-tube-tube-boundary (TTTB) approximation
   for analytical tractability.
The loops we examine are rather long ($L=50~b$), making $a/L$ still rather small
   for the extreme value of $a/b=3$ and hence the {thin-tube (TT)} approximation likely to hold.
However, adopting an $\alpha=5$ means that our transverse profiles are not that steep
   to satisfy the {thin-boundary (TB)} approximation.
Note also that R03 performed an eigen-mode analysis, meaning that
   the damping times therein apply to the exponential stage. 
In contrast, we approach the problem from an initial-value-problem perspective,
   and are therefore able to address whether the damping envelopes of
   the oscillatory signals are purely exponential. 
The net result is that, it is not that informative to quantitatively
   contrast our numerical results (say, the damping times in the exponential stage
   and the periods) with the analytical expectations from R03.
However, a qualitative comparison of the dependence
   of these parameters on $a/b$ and $\rho_{\rm i}/\rho_{\rm e}$ 
   can still be conducted. 

At this point, one may question why do we choose not to examine the 
   linear transverse profile
   in R03 from the outset? 
In fact, we have experimented with this linear profile, and examined the same two density
   ratios but a range of $a/b$ that is only up to $1.4$.
The pertinent results are presented in the appendix, which focuses on steep profiles
   such that the TB approximation holds. 
Basically we find that a Gaussian stage can be seen in addition to an exponential one
   in the damping envelopes.
In addition, a rather close agreement is found between R03 and what we derive for
   the damping times in the exponential stage and the periods. 
On the one hand, this agreement can be seen as a validation of 
   our numerical approach.
On the other hand, it suggests that the qualitative difference between our
   results and R03, if any, is unlikely to result from the small values of 
   plasma $\beta$ in our numerical implementations.
The reason for us not to examine the R03 profile in the main text is
   then twofold. 
First, our profile is as worth examining as the R03 profile, given
   the difficulty to constrain observationally the specific form 
   of the transverse profiles~\citep[e.g.,][and references therein]{2019A&A...622A..44A}.
Second, examining the R03 profile is more computationally expensive. 
As detailed in the appendix, a rather fine grid is needed to resolve the transverse
   inhomogeneity even for an $a/b$ as modest as $1.4$ if we insist on testing
   the TB results. 
The grid resolution is even more demanding for larger values of $a/b$, only for which the
   difference in, say, the periods for modes with different polarizations become 
   substantial.    
If we leave aside the TB consideration, then there will be no analytical expectations
   as happens for the profile given in Equation~\eqref{eq_fr}.

\begin{figure}
\centering
\gridline{\fig{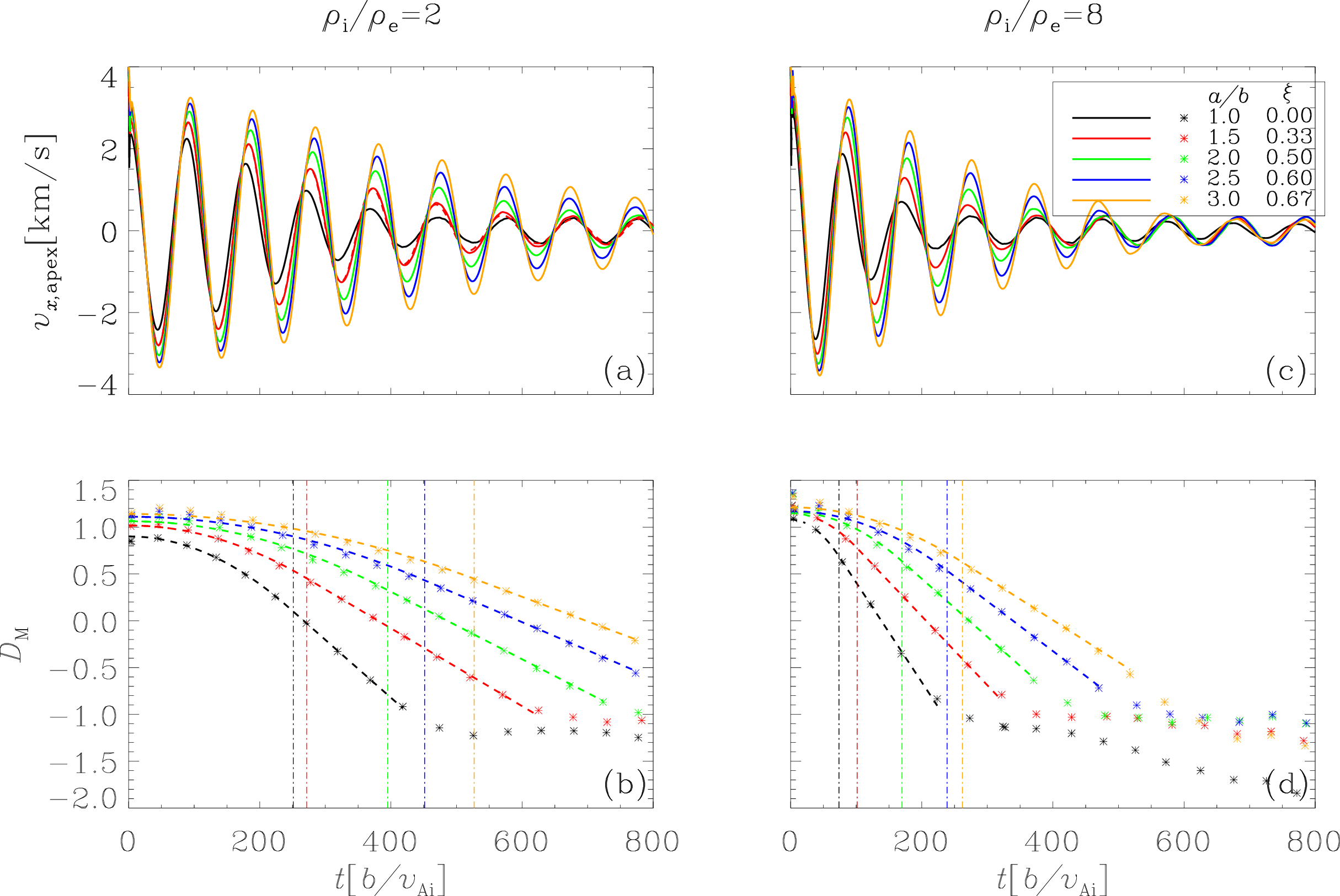}{0.8\textwidth}{}}
\caption{
 Temporal evolution of the $x$-component of the velocity at loop apex
    ($v_{x, {\rm apex}}$) for kink oscillations polarized along the major axis
    (M-modes) supported by loops with elliptic cross-sections and 
    transverse profiles described by Equation~\eqref{eq_fr}. 
Two density ratios ($\rho_{\rm i}/\rho_{\rm e}$) are adopted, 
    one being $2$ (the left column) and the other being $8$ (right).
A number of aspect ratios ($a/b$) are examined as labeled by 
    the different colors. 
In addition to the time series for $v_{x, {\rm apex}}$ itself (the upper row),
    we also extract the extrema therein and plot 
    the natural logarithms of their absolute values ($D_{\rm M}$)
    in the lower row.
The dashed curves represent the best-fit to this discrete series
    with the three-parameter model given in Equation~\eqref{eq_F}, where
    a Gaussian damping envelope is distinguished from an exponential one.
The switch time from one envelope to the other is represented
    by the vertical dash-dotted lines.
Note that the flattening $\xi = 1-b/a$ labelled here represents the deviation of a cross-section from a circular one.          
}			
\label{fig_vx_t}
\end{figure}

Now let us start with our examination of the kink oscillations in coronal loops
   with our prescribed transverse profile
   by looking at the oscillatory behavior of the M-modes. 
The upper row of Figure~\ref{fig_vx_t} presents the temporal
   evolution of the $x$-component of the velocity at
   the loop apex ($v_{x, {\rm apex}} \equiv v_x(0, 0, L/2; t)$) 
   for two density ratios, one being $\rho_{\rm i}/\rho_{\rm e}=2$ (the left column) and the other being $8$ (right).
Different aspect ratios ($a/b$) are differentiated with
   the different colors shown in Figure~\ref{fig_vx_t}b.
From each time series, we extract the extrema $v^{\rm extr}_{x, {\rm apex}}$ and 
   define the damping envelope $D_{\rm M}(t)$ as $\ln |v^{\rm extr}_{x, {\rm apex}}(t)|$.
This discrete $D_{\rm M}$ series is given by the asterisks in the lower row.
We also fit this series with a three-parameter model
\begin{eqnarray}
  D(t)=
\begin{cases}
  \displaystyle
  A^{\rm G}-\frac{1}{2}\left(\frac{t}{\tau^{\rm G}}\right)^2, & t\leq t_{\rm s} \\
  \displaystyle 
  \left[
  A^{\rm G}-\frac{1}{2}\left(\frac{t_{\rm s}}{\tau^{\rm G}}\right)^2
  \right]
    -\left(\frac{t-t_{\rm s}}{\tau^{\rm E}}\right)~,                                              & t> t_{\rm s}
\end{cases}
\label{eq_F}
\end{eqnarray}
    which is motivated by the studies on kink oscillations 
    in loops with circular cross-sections
    \citep[][and references therein]{2016A&A...589A.136P}.
Equation~\eqref{eq_F} separates a Gaussian envelope
    from an exponential one, with the two envelopes characterized by
    the damping times $\tau^{\rm G}$ and $\tau^{\rm E}$, respectively. 
Note that the switch time ($t_{\rm s}$) between the two envelopes is not
    an independent fitting parameter.
Rather, it is defined by $(\tau^{\rm G})^2/\tau^{\rm E}$ {
    as inspired by the studies by \citet{2013A&A...551A..39H} and \citet{2016A&A...589A.136P},
    and has the advantage that $D(t)$ is not only continuous but also smooth
    at $t_{\rm s}$. 
In practice, this means}
    that
    the model indeed involves only three independent parameters.     
The best-fit curves are plotted as the dashed lines
    in the lower row of Figure~\ref{fig_vx_t},
    where we also indicate the derived values of $t_{\rm s}$
    as the vertical dash-dotted lines. 

All time series in the upper row of Figure~\ref{fig_vx_t} start with
    some sudden reduction of the oscillation magnitudes. 
Here by sudden we mean that this phase persists in a time interval that
    is of the order of the transverse rather than the longitudinal Alfv\'en time.
As first found in time-dependent simulations by \citet{2006ApJ...642..533T}, 
    this phase is known as the impulsive leaky phase, during which
    part of the energy imparted by the initial perturbation
    is emitted into the ambient medium.
In other words, the decrease in the oscillation magnitude is connected
    to the energy-confinement capability of coronal loops, which in turn
    depends on the extent to which a loop feels the existence of its ambient medium.
Comparing the curves in different colors in either Figure~\ref{fig_vx_t}b
    or Figure~\ref{fig_vx_t}d, one sees that for
    a given density ratio ($\rho_{\rm i}/\rho_{\rm e}$),
{the magnitude of the oscillations 
    tends to increase
    with the aspect ratio ($a/b$) in the time interval following the impulsive phase
    (see e.g., the asterisks at $t \approx 50~b/v_{\rm Ai}$).}          
This is understandable given that the majority of the velocity
    vectors in the loop region is aligned with the major axis
    (see Figure~\ref{fig_snapshots_Major}).
With $a/b$ increasing, the velocity field becomes increasingly
    tangential to the areas where the loop plasmas are in contact with
    the ambient medium, which decreases the elastic interaction between the loop
    and its surroundings.  
Consequently, the loop becomes less aware of its surrounding medium.  
This heuristic argument, essentially based on inertia considerations,
    also applies to the observation that, 
    for a given $a/b$, the oscillations in the first cycle tend
    to be stronger for a higher density ratio. 
In this case, a loop is less aware of its ambient medium because the ambient
    is more tenuous. 
         
The oscillations rather quickly settle into a periodic behavior after the
    impulsive phase. 
This periodic behavior is reflected by the nearly constant spacing between
    adjacent extrema in the lower row of Figure~\ref{fig_vx_t}, 
    which suggests that a substantial interval can be well fitted with 
    the model given by Equation~\eqref{eq_F}. 
In particular, a Gaussian damping envelope can indeed be distinguished 
    from an exponential one, despite that Equation~\eqref{eq_F}
    was motivated by the results for kink oscillations in
    loops with circular cross-sections.      
However, the oscillations do not weaken indefinitely with time, but rather 
    end up with a rather irregular temporal dependence. 
This may cause our fitting procedure to be problematic
    because if this irregular large-time behavior
    appears too soon, we will either be unable to discern the exponential envelope
    or even have too few extrema to discern any envelope at all. 
While this issue does not arise for the M-modes examined here, it indeed arises 
    for the m-modes to be discussed shortly.     
But before doing that, let us try to further exploit our time-dependent results to 
    shed more light on the temporal evolution 
    of the system. 

Figure~\ref{fig_snapshots_Major} shows a few snapshots of a loop system that experiences
    M-mode oscillations and that corresponds to
    a combination $[a/b, \rho_{\rm i}/\rho_{\rm e}] = [2, 2]$. 
Shown here is the distribution in the transverse { ($xy-$) plane} 
    of the mass density (the filled contours) and the velocity field (arrows) 
    at the loop apex ($z=L/2$). 
Note that the symmetric properties of M-modes enable us to
    show only one half of the system. 
Note also that the density variations are very weak because of the small amplitude 
    of the initial perturbation, and the density contours are simply intended to 
    outline the loop. 
Taken from the animation attached to this figure, the snapshots are intended to show 
    the status of the system at a few representative times.
In particular, Figure~\ref{fig_snapshots_Major}a shows the initial velocity field,  
    which is entirely in the $x$-direction.   
One sees from Figure~\ref{fig_snapshots_Major}b that 
    the impulsive leaky phase is almost immediately followed by the development of
    vortical motions at the outer loop boundary, 
    a characteristic well-known for kink oscillations in
    loops with circular 
    cross-sections~\citep[see e.g., Figure~2 in][]{2014ApJ...788....9G}.  
On the other hand, Figure~\ref{fig_snapshots_Major}c indicates
    that the damping of the M-mode, 
    namely the weakening of the motions in the loop interior,
    is accompanied by the growth of horizontal motions at the loop boundary.
This is a well-known signature of resonant absorption for kink oscillations 
    in loops with circular
    cross-sections~\citep[see e.g., Figure~12 in][]{2014ApJ...788....9G}. 

The same loop system is further examined in Figure~\ref{fig_Major_v_cuts}
    where we plot the distributions of the $x$-component of the fluid velocity
    in the $x=0$ plane (Figure~\ref{fig_Major_v_cuts}a) 
    and the $z=L/2$ plane (i.e., loop apex, Figure~\ref{fig_Major_v_cuts}b)
    at a time when then the system has sufficiently evolved. 
We also show how the distribution of $v_x$ along the $y$-direction
    for $(x, z) = (0, L/2)$ evolves with time in Figure~\ref{fig_Major_v_cuts}c. 
This temporal evolution reinforces the interpretation of the damping
    of the kink oscillation in terms of resonant absorption, given that
    the attenuation of the oscillations in the loop interior ($y \lesssim 0.5~b$)
     takes place in conjunction with the enhancement of $v_x$ around the loop boundary.
Furthermore, one sees that the stripes corresponding to the $v_x$ enhancements      
    become increasingly inclined with time. 
In other words, shorter and shorter spatial scales develop in the $v_x$ profiles, 
    eventually resulting in multiple ripples 
    as seen in Figure~\ref{fig_Major_v_cuts}b.
This behavior is well known for kink oscillations in loops
     with circular cross-sections, and has been customarily interpreted as 
     the phase-mixing of the localized Alfv\'en waves that are resonantly 
     converted from collective kink oscillations
     \citep[see e.g., Figure~12 in][]{2019A&A...631A.105H}.      
What our Figure~\ref{fig_Major_v_cuts} shows is that the resonant coupling of kink modes
     to and the subsequent phase-mixing of localized Alfv\'en waves is robust 
     in that they take place for loops with elliptical cross-sections as well.
It is just that in our case, the ripples are more elongated in the $x$-direction,
     which is understandable given that M-modes are polarized along 
     the major axis. 
As for Figure~\ref{fig_Major_v_cuts}a, one sees the concentration of
     the $v_x$ enhancements at the loop boundary as well. 
These enhancements tend to increase from the loop footpoint
    to loop apex, as expected 
    for fundamental modes.      
{We have performed a more detailed examination on
    the resonant conversion of the kink oscillations into 
    localized Alfv\'enic motions. 
    This examination is presented in Appendix \ref{sec_appendix2}
       to avoid digressing too much from our discussions on the characteristic timescales
       of the decaying kink oscillations.   
}

Now move on to the m-modes. 
Figure~\ref{fig_vy_t} overviews, in a form identical to Figure~\ref{fig_vx_t},
    the temporal evolution of the $y$-component of the fluid velocity 
    at the loop apex ($(x, y, z) = (0, 0, L/2)$). 
Note that the damping profile ($D(t)$) is now assigned with the subscript $m$,
    and derives directly from the extrema in each series of $v_{y, {\rm apex}}$. 
Note also that the curves and symbols corresponding to $a/b = 1$, 
    namely the cases where we examine loops with circular cross-sections,
    are not replotted from Figure~\ref{fig_vx_t} 
    but found with independent simulations. 
However, these curves and symbols coincide exactly with those in 
	 Figure~\ref{fig_vx_t}, despite the different direction of the initial velocity perturbations and the different numerical grid.
This coincidence is reassuring in that for loops with circular cross-sections,
     the M- and m- modes should be degenerate. 
As is the case for Figure~\ref{fig_vx_t}, 
    the impulsive phase can be discerned
    in Figures~\ref{fig_vy_t}a and \ref{fig_vy_t}c, 
    characterized by the initial short-duration reductions
    in the oscillation magnitudes. 
Interestingly, one sees from the lower row that the magnitudes
    {of the oscillations in the time interval
    after the impulsive phase} tend to decrease
    with the aspect ratio $a/b$ for a fixed density ratio
    { (see e.g., the asterisks at $t \approx 40~b/v_{\rm Ai}$)},
    in contrast to what happens for the M-modes. 
This is understandable by invoking also the inertia argument, 
    by which we mean the extent to which a loop is aware of its surrounding fluids.
Now the majority of the fluid parcels in the loop interior
    move along the $y$-direction, meaning that when $a/b$ increases, 
    the velocity field becomes increasing
    normal to the areas where the loop fluids get in contact with their
    surroundings.        
As a consequence, the ambient medium fluids play an increasingly important role
    on the loop oscillations, thereby lowering the capability for loops
    to trap the energy contained in the initial perturbations.      
In view of the inertia argument, one would expect that the oscillations
    in the first cycle strengthen for a larger density ratio,
    and this is indeed seen if one compares
    Figures~\ref{fig_vy_t}b and \ref{fig_vy_t}d.

The m-mode oscillations experience some substantial damping after
    the impulsive phase.      
To show what happens during this interval, we have built 
    an animation in the same format
    as the one attached to Figure~\ref{fig_snapshots_Major} for the same 
    loop examined therein. 
A few representative snapshots are extracted from this animation 
    and presented in Figure~\ref{fig_snapshots_minor}.    
In addition, slightly revising the form of Figure~\ref{fig_Major_v_cuts} 
    to account for the present polarization, 
    we have also constructed the distributions of $v_y$ in 
    appropriate { $xz-$ (Figure~\ref{fig_minor_v_cuts}a), 
    $xy-$ (Figure~\ref{fig_minor_v_cuts}b),  
    and $xt-$ planes} (Figure~\ref{fig_minor_v_cuts}c). 
Similar to the M-modes, one sees the development
    of the characteristic vortical motions after the impulsive leaky phase (Figure~\ref{fig_snapshots_minor}b),
    the accumulation of the Alfv\'en modes 
    at the loop boundary
    (Figures~\ref{fig_snapshots_minor}c), 
    and increasing inclination of the velocity enhancement stripes
    with respect to the $t$-axis 
    (Figure~\ref{fig_minor_v_cuts}c).                     
All these signatures further demonstrate the robustness of 
    the notions of resonant absorption and phase-mixing, ``robust" in the sense
    that these processes operate for loops with circular and elliptic cross-sections alike,
    and for M- and m-modes alike.

Comparing the lower rows in Figures~\ref{fig_vx_t} and \ref{fig_vy_t},
    one sees 
    that for a given combination $[\rho_{\rm i}/\rho_{\rm e}, a/b]$, 
    the m-mode damps more strongly than the corresponding M-mode.
Consequently, the irregular large-time behavior 
    appears earlier and sometimes an exponential envelope
    can no longer be found in our fitting procedure
    (see e.g., the cases with $a/b = 1.5$ and $2$ in Figure~\ref{fig_vy_t}b).
At this point, one may question the origin of this large-time behavior. 
From the computational standpoint, one naturally expects that a damping profile 
    will not diminish indefinitely given the necessarily finite grid resolution,
    even if the profile does approach zero.
We have therefore experimented with a finer grid with a resolution twice higher
    than the base computations for a selected number of m-mode simulations. 
The large-time oscillations indeed reach a lower level, indicating that
    they are at least partly of numerical origin. 
However, we choose not to present the results from the fine-grid computations
    for two reasons.
{ Firstly, they offer no more than one additional extreme for us to perform
    the fitting procedure. 
Secondly, and more importantly, the amplitudes of the oscillations in the
    large-time behavior are no more than a few percent of the magnitude
    of the initial perturbation. 
This makes their observational identification unlikely. }
    
Figure~\ref{fig_Ptau} collects the characteristic timescales 
    for the M- (the circles) and m-modes (diamonds) that we have computed. 
Given from the top to bottom are the periods ($P$), 
    the damping times in the Gaussian stage ($\tau^{\rm G}$)
    and those in the exponential stage ($\tau^{\rm E}$).     
The two density ratios ($\rho_{\rm i}/\rho_{\rm e} = 2$ and $8$)
    are differentiated by the different colors. 
Note that here the timescales are in units of the transverse Alfv\'en time
    ($b/v_{\rm Ai}$), and simply dividing them by $L/b = 50$
    yields the values in units of the longitudinal one ($L/v_{\rm Ai}$).   
Let us recall that the periods are derived by doubling the average
    temporal spacing between 
    two adjacent extrema for any combination $[\rho_{\rm i}/\rho_{\rm e}, a/b]$
    in the interval where the fitting procedure is allowed. 
Let us further recall that the damping times are derived from
    the fitting procedure.          
For the M- (m-) modes, the symbols are connected by the solid (dashed) curves unless 
    the symbols are too sparse.
This happens for the m-modes because one sees
    only a limited number of diamonds in Figure~\ref{fig_Ptau}c,
    which in turn arises due to the absence of an exponential stage.
A small number of the symbols are open, which is intended to suggest that these
    values may not be as reliable as those represented by the solid symbols
    (see e.g., the open diamond corresponding to $a/b=1$ in Figure~\ref{fig_Ptau}b). 
We adopt the following criteria to judge whether 
    a damping time is reliable, which is primarily from the consideration
    of the degree-of-freedom involved in the fitting process. 
Note that while the fitting model (Equation~\ref{eq_F}) nominally involves 
    three independent parameters, only two ($A^{\rm G}$ and $\tau^{\rm G}$)
    are relevant when an exponential stage is absent.
We therefore deem the fitting procedure to be inapplicable altogether
    if there are no more than five (four) extrema that can be fitted 
    when an exponential stage does (does not) exist. 
However, this does not really happen. 
What happens is that when both stages are present,
    sometimes there is only one extreme in either the Gaussian
    or the exponential stage. 
The former situation arises for
    the m-mode and equivalently the M-mode computations
     with $a/b=1$ for a $\rho_{\rm i}/\rho_{\rm e}=8$.
The latter happens for our computation of the m-mode 
    with $a/b=2.5$ and $\rho_{\rm i}/\rho_{\rm e}=8$.
Both situations can be found in Figure~\ref{fig_vy_t}d.  
We consider the values derived in such cases
    to be not that reliable.        

Let us examine the periods first, shown in Figure~\ref{fig_Ptau}a.
For loops with circular cross-sections (the aspect ratio $a/b=1$), 
    one sees the expected behavior that
    the M- and m-modes are degenerate, and their
    periods tend to decrease with the density ratio $\rho_{\rm i}/\rho_{\rm e}$.
This is well known for loops with step ~\citep[e.g.][]{1983SoPh...88..179E}
    and steep transverse profiles~\citep[e.g.,][]{2008A&A...484..851G}.
In the TT limit, the period in both cases reads     
\begin{eqnarray}
\displaystyle 
P = \frac{\sqrt{2}L}{v_{\rm Ai}}\sqrt{1+\frac{\rho_{\rm e}}{\rho_{\rm i}}}~.
\label{eq_circ_TTTB_P}	
\end{eqnarray}    
One further sees that for a given density ratio, the periods of the M-modes (m-modes)
    increase (decrease) with the aspect ratio. 
The same qualitative dependence is seen in the analytical study by R03, 
    despite the TTTB approximation and the different transverse profile adopted therein.
In the zero-$\beta$ calculations, R03 worked out the extreme case
    with $a/b \to \infty$, and Equation~(59) therein shows that
    $P_{\rm M}$ and $P_{\rm m}$ become $2L/v_{\rm Ai}$ and $2L/v_{\rm Ae}$, respectively.
In other words, in the case of infinitely elongated cross-sections, 
    $P_{\rm M}$ does not involve the ambient corona at all,
    whereas $P_{\rm m}$ is entirely determined by the ambient corona.
R03 invoked the inertia argument to explain this extreme behavior, namely
    the polarization of the M-modes (m-modes) means that the elastic interaction
    between the loops and their surroundings vanishes (maximizes). 
As has been discussed for Figures~\ref{fig_vx_t} and \ref{fig_vy_t},
    the same line of thinking applies 
    to the interpretation of 
    the $a/b$-dependence 
    of the oscillation magnitude right after the impulsive leaky phase as well. 
Provided this reasoning is valid, one may expect the following features
    in Figure~\ref{fig_Ptau}, where the periods are measured in units of $b/v_{\rm Ai}$.
For a given $\rho_{\rm i}/\rho_{\rm e}$, with $a/b$ increasing, 
    both $P_{\rm M}$ and $P_{\rm m}$ should eventually
    show little dependence on $a/b$.
Furthermore, at sufficiently large $a/b$, one expects that $P_{\rm M}$ 
    shows little dependence on $\rho_{\rm i}/\rho_{\rm e}$, whereas 
    $P_{\rm m}$ will decrease monotonically with $\rho_{\rm i}/\rho_{\rm e}$.
Both these features are indeed seen in our numerical results.
Combining these features and what happens for $a/b=1$, one can then readily understand
    the overall behavior for the departure of $P_{\rm M}$ from $P_{\rm m}$
    to be stronger for large $a/b$ and large $\rho_{\rm i}/\rho_{\rm e}$.      
    
Now move on to the damping times as shown in Figures~\ref{fig_Ptau}b and \ref{fig_Ptau}c.
For loops with circular cross-sections, \citet{2016A&A...589A.136P} proposed that 
\begin{eqnarray}
\displaystyle 
\frac{\tau^{\rm G}}{P} =
   \frac{2}{\pi   \kappa \epsilon^{1/2}}~,~~~
\frac{\tau^{\rm E}}{P} =   
   \frac{4}{\pi^2 \kappa \epsilon}~,
\label{eq_circ_TTTB_tauGE}  
\end{eqnarray} 
    valid in the TTTB limit and for transverse profiles characterized by a transition layer (TL) linearly connecting a uniform interior to a uniform ambient medium. 
Here by ``linearly" we mean that the mass density depends on the transverse coordinate 
    in a linear fashion.
And by ``TB" we mean that $\epsilon \equiv l/R \ll 1$, where $l$ is the layer width 
    and $R$ is the mean loop radius. 
Note that $P$ in this case is given by Equation~\eqref{eq_circ_TTTB_P},
    and $\kappa$ is determined by the density ratio as 
\begin{eqnarray}
\displaystyle 
\kappa = \frac{\rho_{\rm i}/\rho_{\rm e}-1}{\rho_{\rm i}/\rho_{\rm e}+1}~.
\end{eqnarray} 
Equation~\eqref{eq_circ_TTTB_tauGE} suggests that 
    both $\tau^{\rm G}$ and $\tau^{\rm E}$ decrease with $\rho_{\rm i}/\rho_{\rm e}$.
This qualitative behavior is also seen in our numerical results.
Furthermore, Figures~\ref{fig_Ptau}b and \ref{fig_Ptau}c 
    indicate that the same trend applies 
    to both M- and m-modes at a given $a/b$.
For a given $\rho_{\rm i}/\rho_{\rm e}$, on the other hand, 
    Figure~\ref{fig_Ptau}b suggests that
    $\tau^{\rm G}_{\rm M}$ increases with $a/b$, whereas the opposite happens
    for $\tau^{\rm G}_{\rm m}$.
It is just that $\tau^{\rm G}_{\rm M}$ possesses a stronger $a/b$-dependence than
    $\tau^{\rm G}_{\rm m}$ does. 
The end result is that, for sufficiently large $a/b$, 
    the difference between $\tau^{\rm G}_{\rm M}$ and $\tau^{\rm G}_{\rm m}$
    is significant for both density ratios.        
One further sees from Figure~\ref{fig_Ptau}c that
    $\tau^{\rm E}_{\rm M}$ increases with $a/b$ for a given $\rho_{\rm i}/\rho_{\rm e}$.
However, the dependence of $\tau^{\rm E}_{\rm m}$ on $a/b$ 
    does not seem to be monotonical.
While the limited number of the diamonds does not permit us to draw 
    a firm conclusion on this, it seems nonetheless safe
    to say that $\tau^{\rm E}_{\rm M}$ is longer than $\tau^{\rm E}_{\rm m}$
    for both density ratios. 
Note that on using the analytical expressions given by R03, 
    we have deduced that $\tau^{\rm E}_{\rm M} < \tau^{\rm E}_{\rm m}$
    for $\rho_{\rm i}/\rho_{\rm e} \lesssim 5$ but $\tau^{\rm E}_{\rm M} \gtrsim \tau^{\rm E}_{\rm m}$ when $\rho_{\rm i}/\rho_{\rm e} \gtrsim 5$.   
This former behavior is not seen in Figure~\ref{fig_Ptau}c,
    despite that the latter is.    
This difference from the R03 results may result from the difference in the 
    adopted transverse profiles and/or the questionable applicability of the
    TB approximation given our value of $\alpha$. 
Whatever the source is, this discrepancy strengthens the importance 
    of examining different profile choices and/or moving away from 
    the TB approximation, as far as the properties of kink oscillations
    in loops with elliptic cross-sections are concerned.

\section{Discussion}
\label{sec_discus}
So what seismological applications do we expect, given our computations for 
    kink oscillations in coronal loops with elliptic cross-sections? 
Let us start by noting that aspect ratios ($a/b$) of $2$ or larger 
    are compatible with recent spectroscopic measurements~\citep{2019ApJ...885....7K}.  
Furthermore, 
    observational surveys indicate that 
    the majority of decaying kink oscillations
    are associated with eruptive events in the lower corona
    \citep[e.g.,][]{2015A&A...577A...4Z,2019ApJS..241...31N}.
As such, the initial transverse perturbations
    that impact on a loop 
    are unlikely to prefer the direction of one axis of
    the elliptical cross-section to the direction of the other.  
Adopting the customary assumption that linear theories apply,
    one then expects that the initial perturbations can be decomposed
    into two independent components, one along the major and the other along
    the minor axis.
Consequently, the M- and m-modes can be excited simultaneously. 
On the other hand, it is unlikely that the line of sight (LoS) of an instrument
    is always aligned with a particular axis.
It is therefore possible for both imagers and spectrographs to sample 
    kink oscillations that are actually a superposition of 
    M- and m-modes.          

Our numerical results suggest the following seismological scheme, 
    which is feasible at least in principle. 
For simplicity, let us focus on fundamental kink oscillations,
    and consider only the application of the periods ($P$) and damping times ($\tau$). 
Let us further assume that the plasma $\beta$ for typical active region loops
    is too small to be relevant, and likewise, the loops are long
    enough for the thin-tube approximation to apply. 
It follows from straightforward dimensional analysis that  
    $P$ and $\tau$ can be formally formulated as
\begin{eqnarray}
\displaystyle 
&& P_{\rm M} = \frac{L}{v_{\rm Ai}} \mathcal{P}_{\rm M}
     \left(\rho_{\rm i}/\rho_{\rm e}, a/b, \alpha \right)~,
    			\label{eq_scheme_MajorP} \\
\displaystyle 
&& P_{\rm m} = \frac{L}{v_{\rm Ai}} \mathcal{P}_{\rm m}
     \left(\rho_{\rm i}/\rho_{\rm e}, a/b, \alpha \right)~,
				\label{eq_scheme_MinorP} \\
\displaystyle 
&& \tau^{\rm G}_{\rm M} = \frac{L}{v_{\rm Ai}} \mathcal{G}_{\rm M}
	\left(\rho_{\rm i}/\rho_{\rm e}, a/b, \alpha \right)~,
				\label{eq_scheme_MajorTauG} \\
\displaystyle 
&& \tau^{\rm G}_{\rm m} = \frac{L}{v_{\rm Ai}} \mathcal{G}_{\rm m}
	\left(\rho_{\rm i}/\rho_{\rm e}, a/b, \alpha \right)~,
				\label{eq_scheme_MinorTauG} \\
\displaystyle 
&& \tau^{\rm E}_{\rm M} = \frac{L}{v_{\rm Ai}} \mathcal{E}_{\rm M}
	\left(\rho_{\rm i}/\rho_{\rm e}, a/b, \alpha \right)~,
				\label{eq_scheme_MajorTauE} \\
\displaystyle 
&& \tau^{\rm E}_{\rm m} = \frac{L}{v_{\rm Ai}} \mathcal{E}_{\rm m}
	\left(\rho_{\rm i}/\rho_{\rm e}, a/b, \alpha \right)~.
			\label{eq_scheme_MinorTauE} 
\end{eqnarray}
Evidently, the specific forms of the functions
   $\mathcal{P}$, $\mathcal{G}$, and $\mathcal{E}$ 
   depend on the specific form of the transverse density profile.
While the symbol $\alpha$ is used, it is intended to represent 
   some dimensionless steepness parameter that characterizes the spatial scale 
   of the transverse density distribution
   (say, the ratio of this spatial scale to the half minor-axis). 
In other words, it does not have to have the same meaning as 
   in Equation~\eqref{eq_fr}.
For the profile linear in some elliptic coordinate, 
   R03 has offered explicit forms for $\mathcal{P}_{\rm M}$,
   $\mathcal{P}_{\rm m}$, $\mathcal{E}_{\rm M}$, and $\mathcal{E}_{\rm m}$.
These expressions are applicable provided that the steepness parameter
   ensures the thin-boundary approximation. 
On the other hand, our computations demonstrate that the functions
   in Equations~\eqref{eq_scheme_MajorP} to \eqref{eq_scheme_MinorTauE}
   can be numerically established for the majority of the equilibrium 
   parameters of interest.
Regardless, the point is that the wave quantities involve
   only four equilibrium parameters, namely, 
   the longitudinal Alfv\'en time ($L/v_{\rm Ai}$),
   the density ratio ($\rho_{\rm i}/\rho_{\rm e}$),
   the aspect ratio ($a/b$),
   and the steepness parameter $\alpha$.    
It therefore follows that these four parameters can be inverted for, provided that 
   the M- and m-modes can be adequately separated. 
From the perspective of practical implementation, 
   Figure~\ref{fig_Ptau}a indicates that it is preferable that both the density ratios
   and the aspect ratios are significant.
However, in principle the seismological inversion is not limited to such cases,
   given the enhanced spectral resolution offered by
   modern time-frequency analysis methods
   \citep[e.g.,][]{2018ApJ...856L..16W}.

Is there any indication that the M- and m-modes 
   may have been observed?
For simplicity, let us focus only on the periods, 
   and insist on that only fundamental modes are involved in the observed signals.
We are then inspired to look for multi-periodic signals comprising periodicities
   that are different but remain of the order of the longitudinal Alfv\'en time. 
Such instances have indeed been observed 
    since the early measurements with TRACE 
	(e.g., \citeauthor{2004SoPh..223...77V}~\citeyear{2004SoPh..223...77V},
	\citeauthor{2015ApJ...799..151G}~\citeyear{2015ApJ...799..151G},        
	\citeauthor{2016A&A...593A..53P}~\citeyear{2016A&A...593A..53P},    
	\citeauthor{2019A&A...632A..64D}~\citeyear{2019A&A...632A..64D};
	see also the review by \citeauthor{2009SSRv..149....3A}~\citeyear{2009SSRv..149....3A};
	and references therein).
However, nearly all these observations are spatially resolved, enabling the measurement 
    of the axial distributions of the phases of the component signals.
In general, it was found that     
    the component signal with a long period ($P_{\rm long}$) 
    possesses a different phase profile from 
    the component with a short period ($P_{\rm short}$).
This fact, together with the fact that $P_{\rm long}$ is not far from $2P_{\rm short}$,
    are therefore more in line with the interpretation of the signals
    as comprising a fundamental mode and its axial harmonic.
Under the assumption that coronal loops possess circular cross-sections,     
    this has inspired a series of studies that exploit the period ratio    
    ($P_{\rm long}/P_{\rm short}$) to seismologically deduce
    such equilibrium parameters as 
    the axial density scaleheight~\citep[e.g.,][]{2005ApJ...624L..57A},
    and/or
    the axial magnetic field scaleheight~\citep[e.g.,][]{2008A&A...486.1015V}.
Having said that, we note one exception found in the multi-periodic signal
    associated with path C
    in a loop arcade imaged with TRACE as reported by 
    \citet[][hearafter V04]{2004SoPh..223...77V}.      
In this case two different periods were identified, 
    the shorter one being $P_{\rm short} \approx 315 \pm 144$~sec
    and the longer one being $P_{\rm long} \approx 405 \pm 35$~sec
    (Table II in V04).
Two sets of values are given there, one from a wavelet analysis 
    and the other from a curve-fitting procedure.
We prefer the latter, because it also yields the damping times
    for both periodicities. 
Interestingly, V04 noted that a Gaussian-like damping
    is more appropriate for both periodicities than an exponential damping. 
The derived damping times read $\tau_{\rm short} = 920\pm 290$~sec and 
    $\tau_{\rm long} = 1550\pm 640$~sec for the short- and long-period components, 
    respectively. 
While the amplitude of the long-period signal was found
    with statistical significance to decrease away from
    the loop apex as expected for fundamental modes,
    there was only limited indication for 
    the short-period signal to show the same behavior.
Nonetheless, let us assume that this $P_{\rm long}$ belongs to the M-mode, and attribute 
    $P_{\rm short}$ to the m-mode, with both modes being axial fundamentals.
From Figure~\ref{fig_Ptau}a we see that $P_{\rm M}$ is sensitive to 
    neither $a/b$ nor $\rho_{\rm i}/\rho_{\rm e}$ when $a/b \gtrsim 1.5$.
Dividing the numbers in V04 by a scaling parameter of $\sim 4.3$
    \footnote{
    	For $a/b \gtrsim 1.5$, one sees that $P_{\rm M}\sim 95 b/v_{\rm Ai} = 1.9 L/v_{\rm  Ai}$. Attributing $P_{\rm long}\approx 405$~sec as measured by V04 to this $P_{\rm M}$, we find that the measured periods and damping times need to be scaled by a factor of $405/95 \sim 4.3$ for these values to be placed in Figure~\ref{fig_Ptau}. This is not to say that path C in V04 is required to possess an $L/b$ of $50$ or 
    	a $b/v_{\rm Ai}$ of $4.3$~sec. Rather, what we assume is that the longitudinal Alfv\'en time $L/v_{\rm  Ai}$ attains $405/1.9 \approx 213$~sec, given that $P_{M}$ is expected to be independent of $L/b$ when $L/b$ is sufficiently large. } 
    and placing
    the resulting values in Figure~\ref{fig_Ptau}a,
    one can see that a combination of
    $[a/b, \rho_{\rm i}/\rho_{\rm e}] = [2.5, 2]$ can readily
    account for the observations. 
While admittedly not a proper inversion, 
    this comparison nonetheless shows that the particular measurements associated 
    with a particular loop in V04 are not incompatible with 
    the interpretation in terms of fundamental M- and m-modes
    in loops with elliptic cross-sections. 

Let us accept that decaying kink oscillations are excited by coronal
	eruptive events.
One then expects that the time series of, say, transverse loop displacements
    involving at least two periodicities
    should be rather common to see, if 
    the cross-sections of coronal loops are more likely to be elliptic, or non-circular to be precise. 
This does not seem to be the case, for multi-periodic signals seem
    to be present in only a small fraction of available measurements of kink oscillations.
On top of that, these multi-periodic signals tend to agree more with 
    the understanding in terms of the existence of axial harmonics. 
So why does this ``scarcity of multi-periodic oscillations" happen? 
There are certainly many possible reasons, and we offer but two. 
It may be that when initially displacing coronal loops, the impulsive drivers
    prefer one half axis to the other for some unknown reason. 
Alternatively, it may be that the cross-sections of oscillating loops are not far 
    from a circular one. 
Discussing the second possibility further, we note that oscillating
    loops are only a subset of coronal loops, which in turn occupy 
    only a subset of magnetic fluxtubes~\citep[e.g.,][]{2010ApJ...719.1083S,2013ApJ...775..120M}.
Regarding the cross-sectional properties, if oscillating loops
    are not representative of coronal loops, then
    one natural question arises as to why only those loops with nearly circular 
    cross-sections tend to be displaced.
If, on the other hand, oscillating loops
    are representative of coronal loops, then one naturally questions why
    the density enhancements tend to fill a certain set of magnetic fluxtubes
    in such a manner as to make the cross-sections nearly circular. 
As pointed out by \citet{2013ApJ...775..120M}, if indeed there, 
    then this last selection effect actually places some rather stringent constraint
    on the mechanisms heating the quiescent corona. 
It is highly non-trivial to address any of the above-mentioned issues, given the
    inevitable need to measure the coronal magnetic field. 

The seismological applications of Equations~\eqref{eq_scheme_MajorP} to 
    \eqref{eq_scheme_MinorTauE} have been discussed
    for the ideal situation that both M- and m-modes can be discerned,
    and that the Gaussian and exponential damping
    envelopes can be differentiated. 
We argue that the ``scarcity of multi-periodic oscillations" does not mean that
    these equations cannot be further exploited, for which
    we offer only two possibilities. 
First, it seems necessary to further exploit the catalogs of kink oscillations
    compiled in \citet{2015A&A...577A...4Z}
    and \citet{2019ApJS..241...31N} to look for more multi-periodic signals
    with advanced time-frequency analysis methods. 
Provided that the ideal situation indeed arises, one ends up
    with an inversion problem that is likely to be over-determined, namely, 
    more observables appear than the equilibrium parameters to invert for. 
Second, it is possible to constrain the equilibrium parameters with
    the available measurements even if there is only one periodicity.
Suppose that this periodicity belongs
    to an M-mode, and that a Gaussian envelope can be distinguished from
    an exponential one. 
In this case, Equations~\eqref{eq_scheme_MajorP}, \eqref{eq_scheme_MajorTauG}, 
    and \eqref{eq_scheme_MajorTauE}
    enable the construction of 
    an inversion curve in the
    four-dimensional parameter space spanned by
    $L/v_{\rm Ai}$, $\rho_{\rm i}/\rho_{\rm e}$, $a/b$, and $\alpha$.
While this under-determined inversion seems less than ideal,
    it reaches the same status as what one can do when only
    $P$ and $\tau^{\rm E}$ are exploited for kink oscillations
    in loops with circular
    cross-sections~\citep[e.g.,][]{2007A&A...463..333A,2008A&A...484..851G, 2014ApJ...781..111S}.                           
Whatever the possible uncertainties in $a/b$, this nonetheless 
    offers an independent means for offering the important information
    about the geometrical properties of the cross-sections
    of coronal loops.        

\section{Summary and Concluding Remarks}
\label{sec_summary}

This research was motivated by the apparent lack of a detailed study, from 
   the initial-value-problem perspective, on the decaying kink oscillations
   in coronal loops with elliptic cross-sections. 
A non-circular cross section, however, has been suggested
   on both theoretical and observational grounds.
We approached this problem by numerically following the responses of
   straight loops with elliptic cross-sections to initial transverse velocity perturbations.
We adopted sufficiently weak initial perturbations to focus on kink oscillations
   in the linear regime. 
In addition, inspired by the relevant eigen-mode analysis by
   \citet[][R03]{2003A&A...409..287R},
   two independent sets of initial perturbations were applied, thereby exciting
   kink oscillations with two independent polarizations. 
Based on the pertinent velocity fields, we used M- and m-modes to denote the
   oscillations that are primarily polarized along the major and minor axes, respectively.  
For both polarizations, we performed a rather comprehensive study on 
   how the mode properties depend on the major-to-minor-half-axis ratio $a/b$.
Two internal-to-external density ratios 
   ($\rho_{\rm i}/\rho_{\rm e}=2$ and $8$)
   were adopted to represent the low- and high-density ratio situations.    
Two different specifications of the transverse profiles were also examined. 
One was given by Equation~\eqref{eq_fr} and the pertinent results are presented in the text.
The other was identical to what was adopted by R03 (see Equation~\ref{eq_app_fxy}),
   and the pertinent results are presented in the appendix for validation purposes. 
   
Common to both transverse profiles, and common to both density ratios, 
   we find that the temporal evolution of the transverse displacements 
   is in general characterized by a damping profile that comprises 
   a Gaussian envelope and an exponential one.
In addition, this attenuation of the collective motion is accompanied
   by the enhancement of the transverse Alfv\'enic motions
   in an area surrounding the loop boundary, where small transverse spatial scales
   develop with time. 
In other words, our results lend further support to the robustness of
   the resonant coupling of the kink modes to and the consequent phase-mixing of
   localized Alfv\'enic motions, 
   despite an elliptic rather than a circular loop cross-section.
For all computations we conducted, we find that the periods of the M-modes ($P_{\rm M}$)
   increase with $a/b$ for a given $\rho_{\rm i}/\rho_{\rm e}$,
   whereas the opposite takes place for the periods of the m-modes ($P_{\rm m}$).     
When the aspect ratio $a/b$ is fixed, both $P_{\rm M}$ and $P_{\rm m}$ tend to decrease 
   with the density ratio.
During the Gaussian stage, the characteristic damping time
   for the M-modes $\tau^{\rm G}_{\rm M}$ 
   tends to be longer than that for the m-modes $\tau^{\rm G}_{\rm m}$.
This makes it difficult to find a proper exponential damping envelope
   in some m-mode computations.
For the cases where an exponential envelope exists, 
   the damping time for the m-mode $\tau^{\rm E}_{\rm m}$ may be longer than 
   its M-mode counterpart $\tau^{\rm E}_{\rm M}$ for the profile examined by R03 for
   the low density ratio, even though  
   $\tau^{\rm E}_{\rm m} < \tau^{\rm E}_{\rm M}$    
   in most cases. 

We discussed the potential applications of our numerical findings in 
   the context of coronal seismology, 
   assuming that loops are sufficiently long and the plasma $\beta$ is
   sufficiently small. 
In particular, we showed that one oscillating loop in a loop arcade 
   imaged with TRACE and reported by \citet{2004SoPh..223...77V}
   is likely to host simultaneously an M- and an m-mode.
As such, some information regarding the aspect ratio of the cross-section
   of the oscillating loop can be deduced.
We showed that the inversion problem may be over-determined in the ideal case where 
   an M- and an m- mode are simultaneously measured, and an exponential damping envelope
   can be told apart from a Gaussian one. 
In this case, the number of measureables
   ($P_{\rm M}$, $P_{\rm m}$, 
   $\tau^{\rm G}_{\rm M}$, $\tau^{\rm G}_{\rm m}$,
   $\tau^{\rm E}_{\rm M}$, and $\tau^{\rm E}_{\rm m}$) exceeds the number of 
   the equilibrium parameters to invert for 
   ($L/v_{\rm Ai}$, $\rho_{\rm i}/\rho_{\rm e}$, $a/b$, and $\alpha$).
Here $L/v_{\rm Ai}$ represents the longitudinal Alfv\'en time,
   and $\alpha$ represents some dimensionless steepness parameter
   characterizing the spatial scale of the transverse inhomogeneity. 
It was further shown that some information on $a/b$ can still be gathered
   even in the under-determined cases where  
   only one periodicity can be told, provided that both damping envelopes can
   be measured. 
            
Having said that, we stress that this study is only among the first steps 
   towards addressing kink oscillations in realistic coronal loops. 
For instance, we assumed that the equilibrium parameters are
   homogeneous in the axial direction,
   meaning that we cannot address the influence on the kink mode properties 
   of such effects as the axial density
   stratification~\citep[e.g.,][]{2005ApJ...624L..57A, 2006A&A...457.1059D}
   or loop expansion~\citep[e.g.,][and references therein]{2019A&A...631A.105H}.    
Likewise, neglecting the loop curvature means that we cannot address the
   importance relative to resonant absorption of, say, lateral leakage 
   for damping kink
   oscillations~\citep{2006ApJ...650L..91T, 2008ApJ...682.1338M,2011ApJ...726...42S}.     
Finally, we chose to examine only the linear regime to initiate our studies on 
   kink oscillations in loops with elliptic cross-sections. 
For stronger oscillations, the velocity shear associated with both the kink mode itself
   and the phase-mixed Alfv\'enic motions is known to be more prone to 
   the Kelvin-Helmholtz instability
   \citep[e.g.,][]{1983A&A...117..220H,1984A&A...131..283B, 
   	2015ApJ...809...72A, 
   	2016A&A...595A..81M, 
   	2019ApJ...883...20G,
   	2019MNRAS.482.1143H}. 
It will be informative to examine what happens if we replace the circular cross-sections
    in this extensive series of studies with an elliptic one.

\acknowledgments
This research was supported by the 
    National Natural Science Foundation of China (BL: 41674172, 11761141002, 41974200),
    and also by the European Research Council (ERC) under the European Union’s Horizon
    2020 research and innovation program (TVD via grant agreement No.724326). 
MZG gratefully acknowledges the support from the China Scholarship Council (CSC).

\clearpage
\appendix

\section{Kink oscillations in coronal loops with elliptic cross-sections: 
	A validation study}
\label{sec_appendix}

While the numerical implementation of 
      both our equilibrium setup and solution procedure
      seems rather straightforward (section~\ref{sec_NumModel}),
      it remains reassuring if this implementation can be validated against available
      analytical results. 
However, no known theories exist to our knowledge that pertain to a transverse 
      profile realized through Equation~\eqref{eq_fr}.
In fact, the only analytical study that addresses 
      the resonant absorption of kink oscillations in coronal loops with 
      elliptic cross-sections seems to be the one by
      \citet[][hereafter R03]{2003A&A...409..287R}.              
This section is therefore intended to replace the function $f(x,y)$ with 
      the one examined by R03 and see whether the theoretical expectations
      therein can be reproduced. 
      
Before proceeding, it is necessary to recall the assumptions made by R03
    and the main results therein.
Basically, R03 employed zero-$\beta$ MHD throughout, and 
    adopted the thin-tube-thin-boundary (TTTB) approximation.
Furthermore, R03 worked in an elliptic coordinate
    system $(s, \varphi)$ in the transverse plane $(x, y)$,
\begin{eqnarray}
  x = \sigma \cosh s \cos\varphi~,~~~
  y = \sigma \sinh s \sin\varphi~,
\label{eq_coorTrans_ellip}  
\end{eqnarray} 
    where $s$ ranges from $0$ to $\infty$ and $\varphi$ varies 
    between $-\pi$ and $\pi$.
Any contour of the coordinate $s$ is an ellipsis, and R03 chose some $s_0$
    to denote the outer loop boundary beyond which the medium is homogeneous.
Letting $a$ and $b$ denote the major and minor half-axes associated with this boundary
    ($a > b$), 
    one finds that $\sigma$ and $s_0$ are then determined through
\begin{eqnarray}
\displaystyle 
    \sigma = \sqrt{a^2-b^2}~, ~~~
    s_0    = \ln\frac{a+b}{\sqrt{a^2-b^2}}~.
\label{eq_app_sig_s0}    
\end{eqnarray}  
The TT approximation translates to that $a/L \ll 1$ where
    $L$ is the loop length.  
R03 then assumed that the density varies in a form identical to Equation~\eqref{eq_rhoEq},
    with $f(x, y)$ now reading
\begin{eqnarray}
f(x, y) \equiv f(s) = 
  \left\{
  \begin{array}{ll} 
   1~,                    &                 s < s_0-\delta~, \\
   (s_0-s)/\delta~,       & s_0-\delta \leq s \leq s_0~, \\
   0~,                    &                 s> s_0~.
  \end{array} 
  \right. 
\label{eq_app_fxy}
\end{eqnarray}
Here a transition layer (TL) is sandwiched between
   a uniform interior and a uniform ambient medium.
In the elliptic coordinate system, the width of this TL is $\delta$ in the $s$-direction, 
   regardless of $\varphi$.
However, when translated to a Cartesian coordinate system, the width varies from $b \delta$
   in the $x$-direction to $a\delta$ in the $y$-direction. 
By TB, R03 means that $\delta/s_0 \ll 1$. 

The following results were then obtained by R03.
Focusing on fundamental modes, namely the kink modes with an axial wavenumber 
    $k = \pi/L$, Equation~(60) in R03 leads to that the periods ($P$) are given by
\begin{eqnarray}
  P_{\rm M} 
= \frac{2L}{v_{\rm Ai}}
  \sqrt{\frac{\zeta\rho_{\rm ie}+1}{\rho_{\rm ie}\left(\zeta+1\right)}}~,~~
  P_{\rm m} 
= \frac{2L}{v_{\rm Ai}}
  \sqrt{\frac{\rho_{\rm ie}+\zeta}{\rho_{\rm ie}\left(\zeta+1\right)}}~.
\label{eq_app_P}
\end{eqnarray}
Here $v_{\rm Ai}$ is the Alfv\'en speed at the loop axis,
  and we have used the shorthand notations 
  $\zeta$ and $\rho_{\rm ie}$ to represent the aspect ratio $a/b$
  and density ratio $\rho_{\rm i}/\rho_{\rm e}$, respectively.
In addition, the subscripts M and m correspond to the M- and m-modes, 
  namely the modes polarized along
  the major and minor axes, respectively.   
For both modes, we obtain the damping times ($\tau$) by plugging
   Equations~(84) and (85) into (78) and (79) in R03, the results being
\begin{eqnarray}
   \tau_{\rm M} 
 = \left(1+\frac{1}{\zeta}\right)^2
   \frac{\zeta\rho_{\rm ie}+1}{\rho_{\rm ie}-1}
   \frac{P_{\rm M}}{\pi^2\delta}~,~~ 
   \tau_{\rm m} 
 = \left(1+\zeta\right)^2
   \frac{\rho_{\rm ie}+\zeta}{\zeta\left(\rho_{\rm ie}-1\right)}
   \frac{P_{\rm m}}{\pi^2\delta}~.
\label{eq_app_tau}
\end{eqnarray}

Now the question is whether the analytical results from the eigen-mode analysis in R03
    can be reproduced by time-dependent 3D MHD simulations.
Numerically speaking, almost everything, be it the setup of the numerical grid system
    or the way for exciting kink oscillations, can be inherited
    from Section~\ref{sec_NumModel}.
It is just that $f(x,y)$ in Equation~\eqref{eq_fr} needs to be replaced with 
    Equation~\eqref{eq_app_fxy}.    
Consequently, the steepness parameter $\alpha$ therein is now replaced with $\delta$.
As in the main text, we fix $b$ to be $1000$~km, but allow $a/b$ to vary. 
However, one sees from Equation~\eqref{eq_app_sig_s0} that $s_0$ decreases with
    increasing $a/b$.
This means that when some $\delta$ is given, $\delta/s_0 \ll 1$
    is not necessarily guaranteed.      
Take $\delta = 0.2$ for instance.
For an $a/b$ of $1.04$, one finds that $s_0 = 1.97$, and $\delta/s_0$ indeed 
    satisfies the nominal criteria (say, $\delta/s_0 \le 0.1$) for $\delta/s_0$
    to be considered small. 
However, if one takes $a/b = 1.5$, then $s_0 = 0.81$ and hence $\delta/s_0$ attains
    $0.25$, a value that is no longer much smaller than unity. 
Therefore in practice, we fix $\delta/s_0$ to be $0.1$ for validation purposes
    while varying $a/b$ 
    between $1.1$ and $1.4$.
Note that we avoid the limit $a/b = 1$ because
    of the appearance of $\sqrt{a^2-b^2}$ in
    Equation~\eqref{eq_app_sig_s0}.
On the other hand, $s_0$ decreases with $a/b$, meaning that $\delta$
    and hence the smallest TL width
    ($b\delta$ along the direction of the major axis) decrease as well.            
In fact, this poses a computational issue because a cell size as small as $5$~km
    is necessary for us to resolve the TL for an $a/b$ as modest as $1.4$.
We therefore choose not to increase $a/b$ any more.    

In addition to varying $a/b$, we will also experiment with two choices of
   the density ratio $\rho_{\rm i}/\rho_{\rm e}$, one being $2$ 
   and the other being $8$.
These two choices are not arbitrary.
Rather, they are intended to bring out the different behavior of the damping times
   ($\tau$) as expected
   with Equation~\eqref{eq_app_tau}.
Note that Equation~\eqref{eq_app_P} suggests that, at any given $a/b >1$, the period
  of the M-mode ($P_{\rm M}$) always exceeds that of the m-mode ($P_{\rm m}$) for
  any $\rho_{\rm i}/\rho_{\rm e}$.
However, one may readily show that when $a/b$ is given, $\tau_{\rm m}$ is
   longer than $\tau_{\rm M}$ when $\rho_{\rm i}/\rho_{\rm e} \lesssim 5$,
   whereas the opposite is true when $\rho_{\rm i}/\rho_{\rm e} \gtrsim 5$.
To see this, we note that Equation~\eqref{eq_app_tau} leads to that
\begin{eqnarray}
\displaystyle 
  \frac{\tau_{\rm m}}{\tau_{\rm M}}
= \left(\frac{a}{b}\right) 
  \left[
       \frac{1+(a/b)(\rho_{\rm e}/\rho_{\rm i})}{(a/b)+(\rho_{\rm e}/\rho_{\rm i})}
  \right]^{3/2}~~.
\label{eq_app_taum_O_tauM} 
\end{eqnarray} 
Suppose for now $a/b$ exceeds unity by only a small amount $\varepsilon$
   ($0 < \varepsilon \ll 1$).
Taylor-expanding the right-hand side of Equation~\eqref{eq_app_taum_O_tauM}
   and retaining only terms that are first-order in $\varepsilon$, one finds that 
\begin{eqnarray}
\displaystyle 
  \frac{\tau_{\rm m}}{\tau_{\rm M}}
\approx 
  1+\varepsilon \frac{5-\rho_{\rm i}/\rho_{\rm e}}{2+\rho_{\rm i}/\rho_{\rm e}}~.
\label{eq_app_taum_O_tauM_taylor} 
\end{eqnarray} 
If then follows that the behavior of $\tau_{\rm m}/\tau_{\rm M}$ changes when
   $\rho_{\rm i}/\rho_{\rm e}$ exceeds some critical value
   $(\rho_{\rm i}/\rho_{\rm e})_{\rm c}$, which reads $5$ for 
   $a/b$ not substantially different from unity.
Some further numerical evaluation demonstrates that    
   $(\rho_{\rm i}/\rho_{\rm e})_{\rm c}$ weakly increases with $a/b$,
   reading $5.7$ for an $a/b$ as large as $3$.

What difference do we expect regarding our numerical results relative
    to the analytical ones given by R03, accepting that $\delta/s_0 = 0.1$
    is sufficiently small to satisfy the TB approximation?
This should come from three sources. 
First,  the plasma $\beta$ in our simulations is not zero, whereas zero-$\beta$ MHD
    is adopted by R03 from the outset. 
Nonetheless, the largest $\beta$, a fixed value attained at the loop axis
    ($\beta_{\rm i}$), 
    is merely $0.043$.      
Second, the largest $a/L$ we examine will be $1.5b/L = 1.5/50$, 
    which seems to be sufficiently small.
Third, we will derive the periods ($P$) and damping times ($\tau$)
    largely by
    applying a fitting procedure to the relevant time series. 
As such, some uncertainty may arise due to the grid resolution,
    and  
    the temporal spacing between the output data slices.
However, we have experimented with different choices on these aspects to ensure
    that the derived values for $P$ and $\tau$ converge.
All in all, this means that these derived values can be considered to agree 
    with the expectations from R03 if the relative difference
    is of the order of $\beta_{\rm i}$.

We start with a description of the M-modes.
The upper row of Figure~\ref{fig_app_M}
    presents the temporal evolution of the $x$-component of the velocity at
    the loop apex ($v_{x, {\rm apex}} \equiv v_x(0, 0, L/2; t)$) 
    for two density ratios, one being $\rho_{\rm i}/\rho_{\rm e}=2$ (the left column) and the other being $8$ (right).
A number of aspect ratios are examined, as represented by the different colors
    shown in Figure~\ref{fig_app_M}a. 
From the time series for each combination of $[a/b, \rho_{\rm i}/\rho_{\rm e}]$, 
    we extract the extrema $v^{\rm extr}_{x, {\rm apex}}$ and 
    define the damping envelope ($D_{\rm M}$)
    as the natural logarithms of their absolute values.
This discrete series, represented by the asterisks, 
    is fitted with the three-parameter model as given by 
    Equation~\eqref{eq_F} in the text to distinguish a Gaussian stage 
    from an exponential one. 
The best-fit model is then plotted with a dashed curve.
In addition, the switch time $t_{\rm s}$ between these two stages
    is represented by the vertical dash-dotted line. 
The extrema, best-fit curves, and switch times are all given in the lower row
    and color-coded according to $a/b$
    in a way consistent with the upper row. 
When performing the fitting procedure, we exclude the extrema
    for the initial time interval of the order of
    the transverse Alfv\'en time.
As explained in the text, this time interval
    is likely to be connected with the impulsive leaky phase 
    and therefore irrelevant for our further analyses
    \citep{2006ApJ...642..533T}.    
         
Figure~\ref{fig_app_m} presents, in a form identical to Figure~\ref{fig_app_M}, the relevant results
    for the m-modes. 
A cursory comparison between Figures~\ref{fig_app_M} and \ref{fig_app_m}, the lower rows in particular,   
    indicates that the periods and damping times
    are different for the M- and m-modes, and this difference depends
    on the density ratio.
This difference will be examined shortly.
For now it suffices to note that be it the M- or the m-modes, the impulsive leaky
    phase is less clear for the larger density ratio, resulting in 
    an oscillation of a larger magnitude in the first couple of cycles.
This is expected, given that a larger density ratio makes the coronal loop more 
    efficient in trapping the energy imparted by the initial perturbation.
More importantly, the lower rows of Figures~\ref{fig_app_M} and \ref{fig_app_m} indicate that
    a Gaussian stage can in general be told apart from an exponential one as happens for the profile examined in the text,
    despite that a different transverse density distribution is
    adopted.       
     
Figure~\ref{fig_app_R03} summarizes our examination of the kink oscillations pertinent to the 
    density profile adopted by R03.
Here as in the text, the periods are found by first averaging
    the temporal spacing between two adjacent extrema in the interval
    where we perform the fitting procedure, and then multiplying
    this average by two. 
These periods ($P$, Figure~\ref{fig_app_R03}a), together with
    the damping times in the Gaussian stage ($\tau^{\rm G}$, Figure~\ref{fig_app_R03}b)
    and the exponential stage ($\tau^{\rm E}$, Figure~\ref{fig_app_R03}c),
    are plotted against $a/b$ as circles (diamonds) for 
    the M- (m-) modes. 
In addition, the analytical results expected by R03
    (Equations~\ref{eq_app_P} and \ref{eq_app_tau})
    are given by the solid (dashed) curves for the M- (m-) modes as well.
Given the lack of analytical results for the damping times in 
    the Gaussian stage, no curves appear in Figure~\ref{fig_app_R03}b.    
Different colors are adopted for different density ratios, as shown
    in Figure~\ref{fig_app_R03}c. 
Note that the curves for $\tau^{\rm E}$ start with an $a/b$ of $1.1$
    rather than $1$.
This is not to say that, in the limit of $a/b \rightarrow 1$, Equation~\eqref{eq_app_tau} does not converge
    to the well-known TTTB result for loops with circular cross-sections
    (given by e.g., Equation~6 in 
    \citeauthor{2002A&A...394L..39G}~\citeyear{2002A&A...394L..39G}).
Rather, that TTTB result can be readily recovered if we insist
    on adopting an $a/b$-independent $\delta$, in which case 
    $\delta$ translates into the ratio between the TL width
    and the loop radius for a circular cross-section. 
When one chooses to fix $\delta/s_0$, as we do due to computational concerns, 
    $\delta$ diverges because $s_0$ diverges when $a/b$ approaches unity. 
We choose not to examine this situation.  
On the other hand, $\delta$ does not appear in the expressions
    for $P$ in the TTTB limit, allowing us to plot $P$
    for the entire range of $a/b$.

Comparing the curves and symbols in Figures~\ref{fig_app_R03}a and \ref{fig_app_R03}c, 
    one sees that the analytical results in R03 are quantitatively
    reproduced.
Regarding the periods ($P$), our numerical results agree with R03 in
    reproducing the tendency for $P_{\rm M}$ ($P_{\rm m}$) to increase
    (decrease) with $a/b$, and the tendency for the departure of $P_{\rm M}$
    from $P_{\rm m}$ to be stronger for a 
    larger $\rho_{\rm i}/\rho_{\rm e}$.    
Regarding the damping times in the exponential stage ($\tau^{\rm E}$),
    one sees that the analytically
    expected behavior for both $\tau^{\rm E}_{\rm M}$  
    and $\tau^{\rm E}_{\rm m}$
    to increase with $a/b$ is reproduced.
Note that this behavior is not to be confused with Figure~5 in R03.
When translating the damping rate into the damping time, that figure
    suggests that while $\tau^{\rm E}_{\rm M}$ tends to
    increase monotonically with $a/b$,
    $\tau^{\rm E}_{\rm m}$ depends on $a/b$ in a nonmonotonical manner. 
The reason for this difference is once again our choice of fixing $\delta/s_0$
    rather than $\delta$, which in turn is due to concerns of computational cost.
With $\delta/s_0$ fixed and $s_0$ decreasing with $a/b$, the dependence on $a/b$
    in our Figure~\ref{fig_app_R03}c is strengthened relative to Figure~5 in R03.     
Regardless of the way of specifying $\delta$, the curves in Figure~\ref{fig_app_R03}c
    indicate that $\tau^{\rm E}_{\rm M} < \tau^{\rm E}_{\rm m}$
    for a $\rho_{\rm i}/\rho_{\rm e}$ of $2$, 
    whereas this tendency is reversed when $\rho_{\rm i}/\rho_{\rm e} = 8$.    
This is expected with Equation~\eqref{eq_app_taum_O_tauM_taylor}. 
It is just that $\tau^{\rm E}_{\rm M}$ is very close
    to $\tau^{\rm E}_{\rm m}$ for a $\rho_{\rm i}/\rho_{\rm e}$ of $8$.
This latter behavior is also expected, for Equation~\eqref{eq_app_taum_O_tauM}
    suggests that $\tau^{\rm E}_{\rm m}/\tau^{\rm E}_{\rm M}$
    becomes essentially independent on $\rho_{\rm i}/\rho_{\rm e}$
    when $\rho_{\rm i}/\rho_{\rm e}$ becomes sufficiently large.  
Despite the minor difference between the symbols 
    and the curves in blue, our numerical results nonetheless
    capture this $\rho_{\rm i}/\rho_{\rm e}$-insensitivity
    for a sufficiently large $\rho_{\rm i}/\rho_{\rm e}$. 
Without the necessary analytical results to compare with,     
    the symbols in Figure~\ref{fig_app_R03}b are given to simply
    restate that a Gaussian damping envelope can be found in addition 
    to an exponential one even with the profile adopted by R03.
Furthermore, one thing irrespective of how $\delta$ is specified
    is that the values for $\tau^{\rm G}$ for the M-modes tend to exceed 
    the ones for the m-modes for both density ratios, which is 
    at variance with the behavior of $\tau^{\rm E}$.

\section{Resonant Conversion of Kink Oscillations into Alfv\'en Modes}
\label{sec_appendix2}    
\begin{figure}
	\centering
	\gridline{\fig{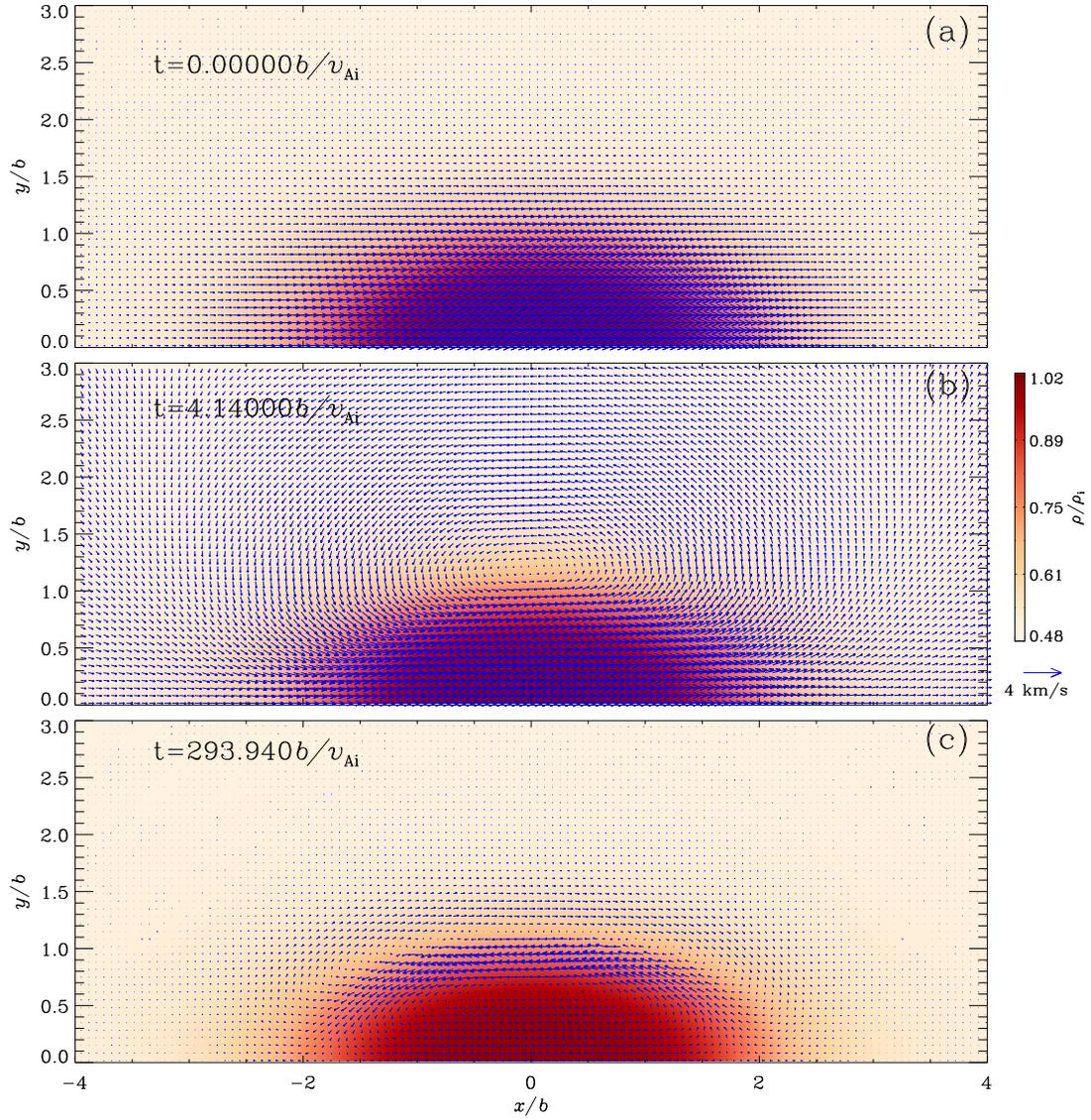}{0.8\textwidth}{}
	}
	\caption{
		Distributions in the transverse ($x-y$) plane of
		    the mass density ($\rho$, the filled contours)
		    and the velocity field (arrows) at the loop apex ($z=L/2$)
		    at a selected number of times as labeled.
		Examined here is a loop that experiences kink oscillations
		    polarized along the major axis, 
		    and therefore only one half of the system is shown.          
        This loop is characterized by 
            an aspect ratio $a/b$ of $2$ and a density
            ratio $\rho_{\rm i}/\rho_{\rm e}$ of $2$.
		The snapshots are taken from the animation attached to this figure. 
	}			
	\label{fig_snapshots_Major}
\end{figure}

\begin{figure}
	\centering
	\gridline{\fig{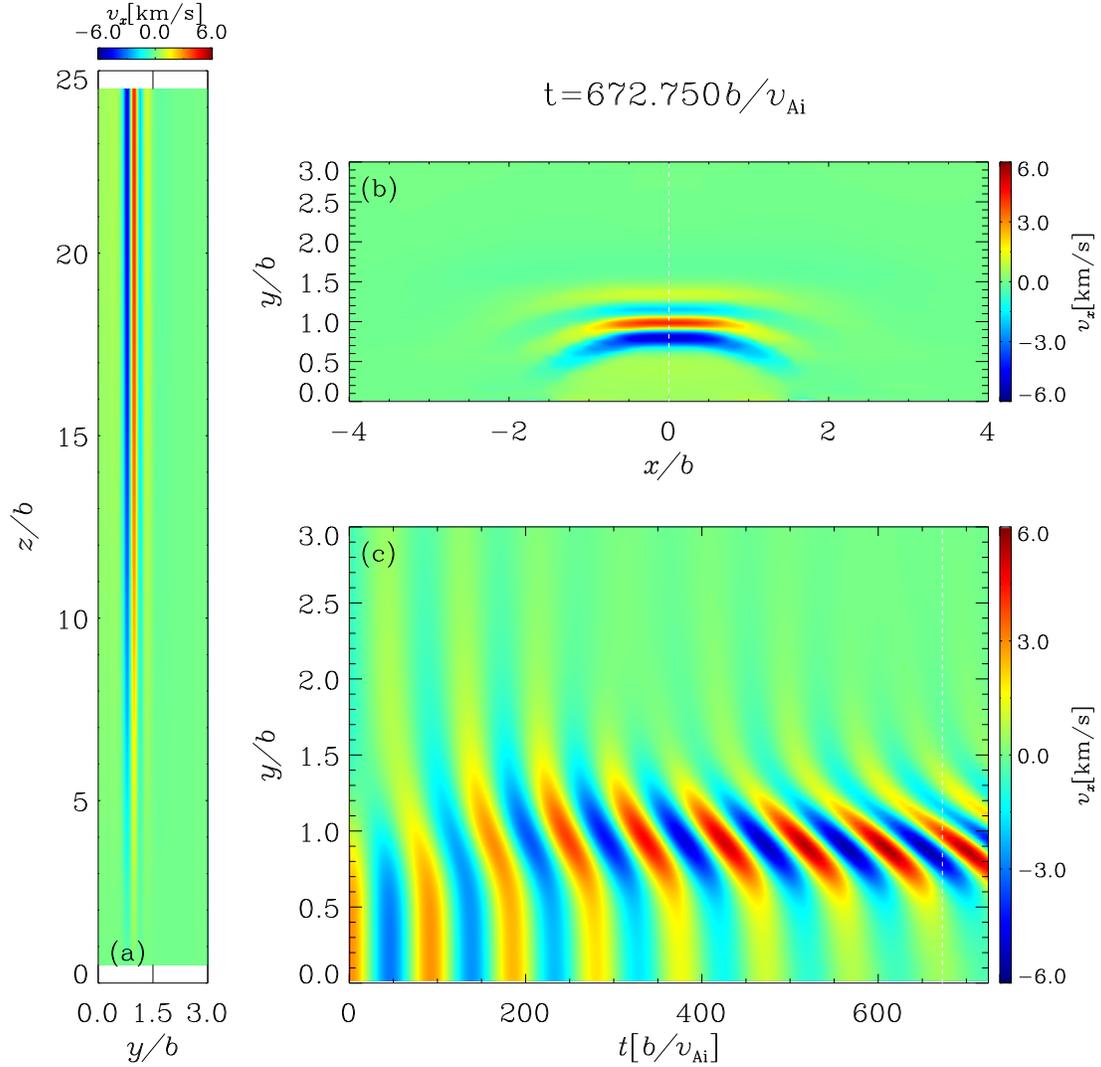}{0.8\textwidth}{}
	}
	\caption{
		Distributions of the $x$-component of the fluid velocity ($v_x$)
		    in (a) the $x=0$ plane and (b) the $z=L/2$ plane
		    at a time when the system has sufficiently evolved.
        Shown in (c) is the temporal evolution of the distribution of $v_x$
            along the $y$ direction for $(x, z) = (0, L/2)$. 
		Examined here is a loop that experiences kink oscillations
			polarized along the major axis.          
		This loop is characterized by 
			an aspect ratio $a/b$ of $2$ and a density
			ratio $\rho_{\rm i}/\rho_{\rm e}$ of $2$.
	}			
	\label{fig_Major_v_cuts}
\end{figure}

\begin{figure}
\centering
\gridline{\fig{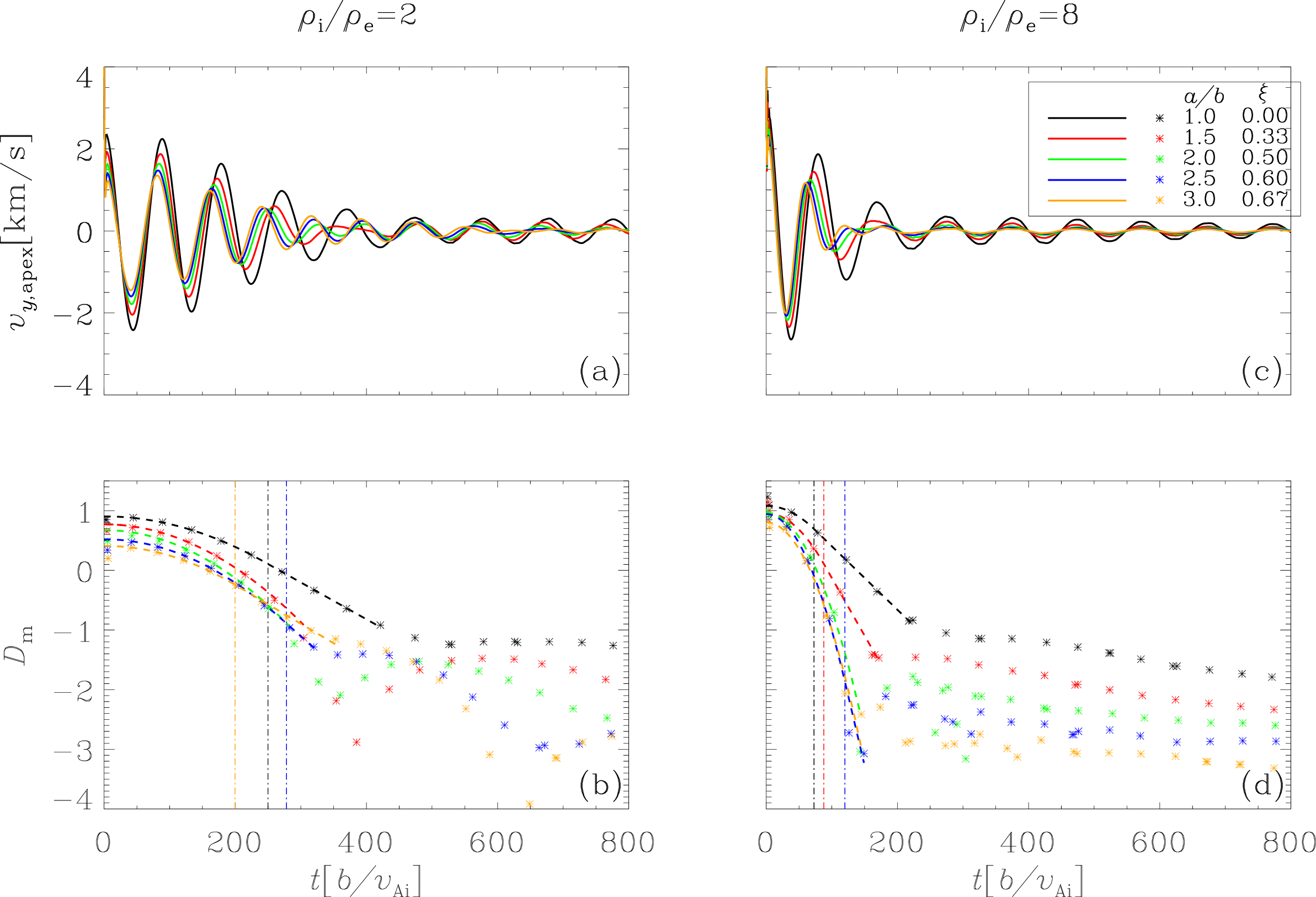}{0.8\textwidth}{}
	}
\caption{
  Similar to Figure \ref{fig_vx_t}, but for kink oscillations polarized
      along the minor axis (m-modes).
  And hence the subscripts $y$ in $v_{y, {\rm apex}}$ and m in $D_{\rm m}$.  
}			
\label{fig_vy_t}
\end{figure}

\begin{figure}
	\centering
	\gridline{\fig{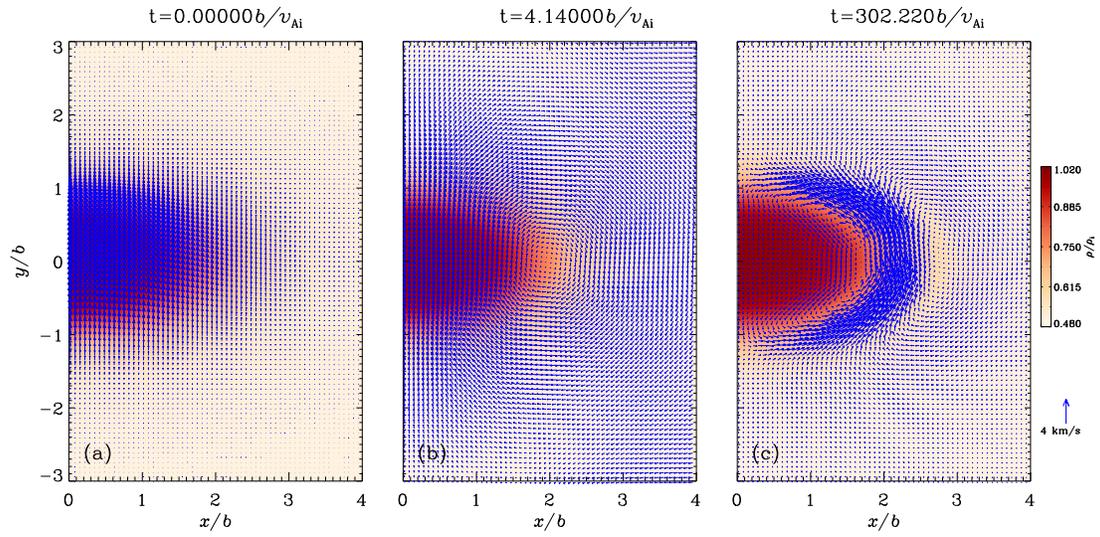}{0.8\textwidth}{}
	}
	\caption{
		Similar to Figure \ref{fig_snapshots_Major}, but for kink oscillations polarized
		along the minor axis.
	}			
	\label{fig_snapshots_minor}
\end{figure}

\begin{figure}
	\centering
	\gridline{\fig{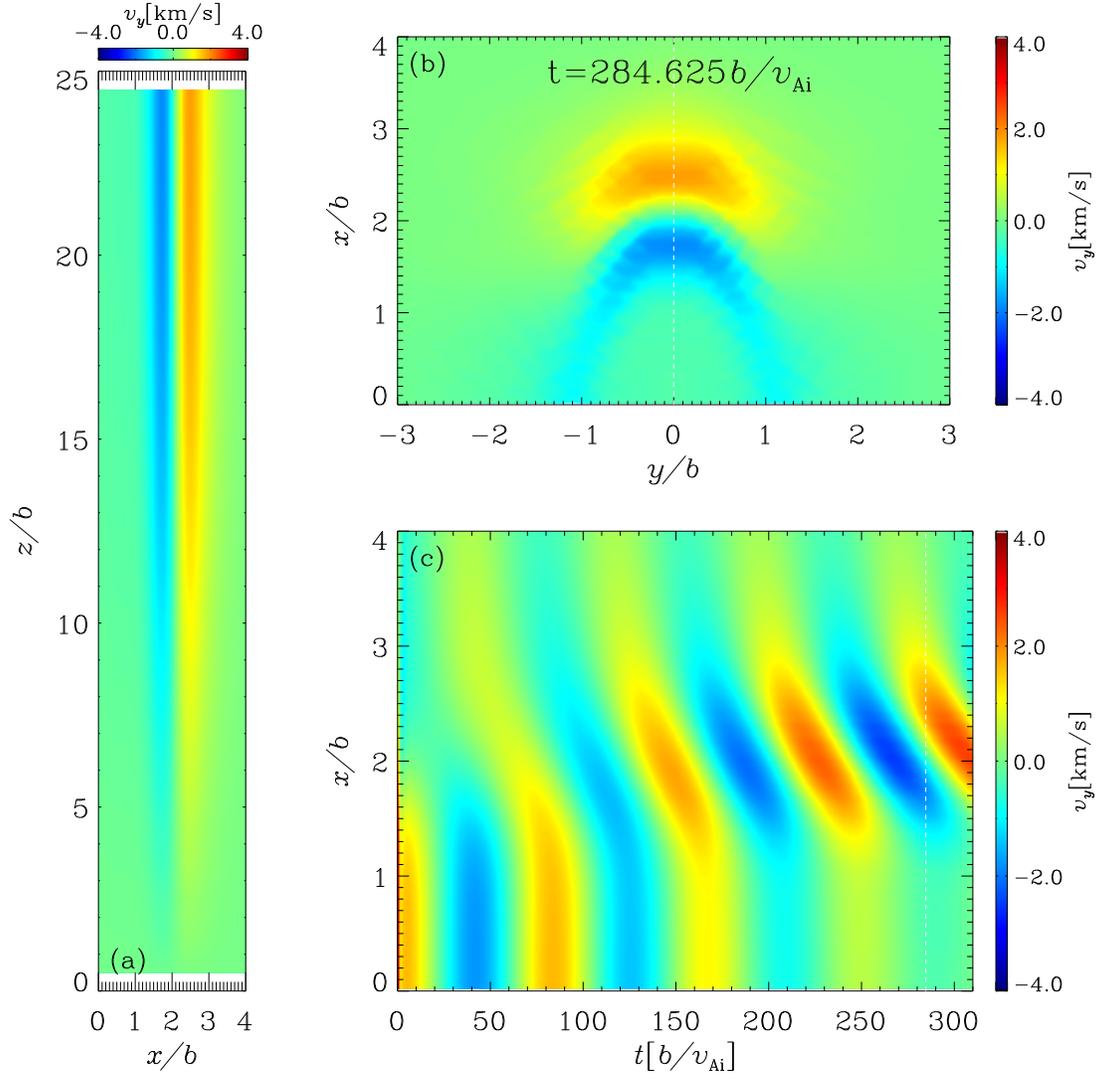}{0.8\textwidth}{}
	}
	\caption{
		Similar to Figure \ref{fig_Major_v_cuts}, but for kink oscillations polarized
			along the minor axis.
		Presented here are the distributions of $v_y$ 
		    in (a) the $y=0$ plane and (b) the $z=L/2$ plane
            at a time when the system has sufficiently evolved.
        Furthermore, given in (c) is the temporal evolution
            of the $x$-distribution of $v_y$
            for $(y, z) = (0, L/2)$. 
	}			
	\label{fig_minor_v_cuts}
\end{figure}

\begin{figure}
\centering
\gridline{\fig{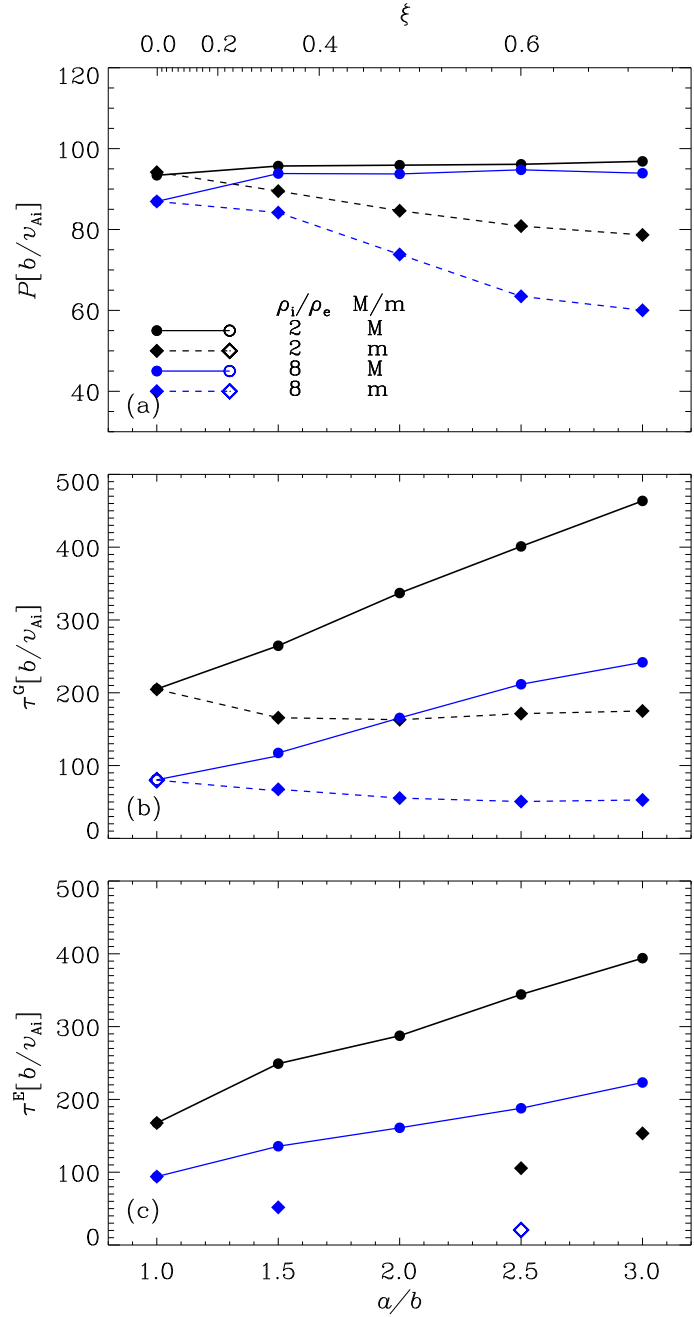}{0.5\textwidth}{}
	}
\caption{
  Dependence on the aspect ratio ($a/b$) of 
    (a) the periods ($P$)
    and (b, c) damping times ($\tau^{\rm G}, \tau^{\rm E}$)
    of the kink oscillations supported by loops with elliptic cross-sections and 
    transverse profiles described by Equation~\eqref{eq_fr}. 
Two different density ratios are examined as labeled by 
    the different colors. 
The symbols represent the values found via best-fitting
    our numerical results, and the open symbols represent the values
    that are not as reliable as the solid symbols.
The circles (diamonds) correspond to
    the M- (m-) modes, namely the modes polarized along the major (minor) axis.    
When not too sparse, the symbols are connected by the solid and dashed curves,
    respectively.  
The flattening $\xi = 1-b/a$ labelled here represents the deviation of a cross-section from a circular one.   
}			
\label{fig_Ptau}
\end{figure}

\begin{figure}
\centering
\gridline{\fig{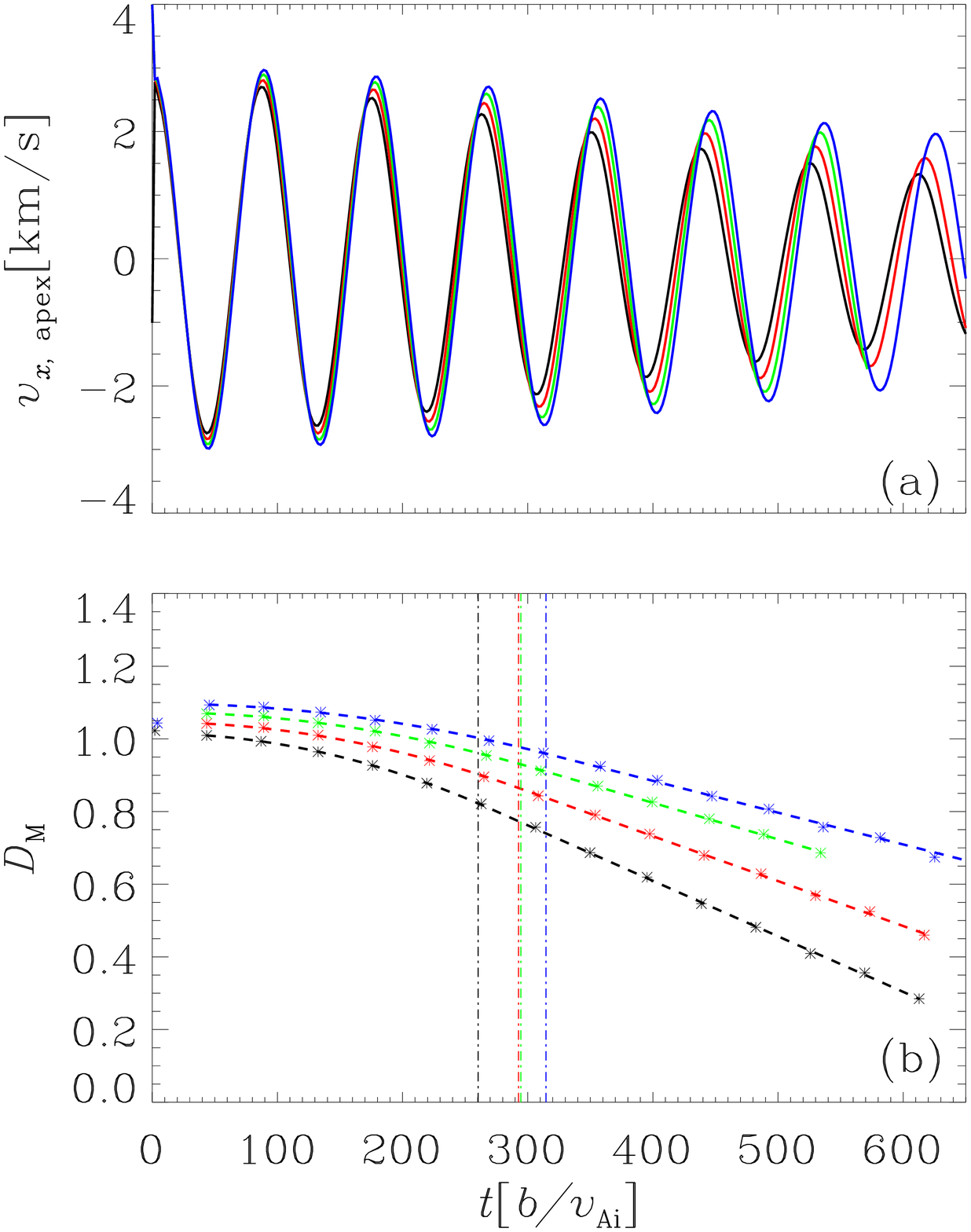}{0.85\textwidth}{}}
\caption{
 Temporal evolution of the $x$-component of the velocity at loop apex
     ($v_{x, {\rm apex}}$) for kink oscillations polarized along the major axis
     (M-modes) supported by loops with elliptic cross-sections and 
     linear transverse profiles described by Equation~\eqref{eq_app_fxy}. 
 Two density ratios ($\rho_{\rm i}/\rho_{\rm e}$) are adopted, one being
     $2$ (the left column) and the other being $8$ (right).
 A number of aspect ratios ($a/b$) are examined as labeled by 
     the different colors. 
 In addition to the time series for $v_{x, {\rm apex}}$ itself (the upper row),
     we also extract the extrema therein and plot 
     the natural logarithms of their absolute values ($D_{\rm M}$)
     in the lower row.
 The dashed curves represent the best-fit to this discrete series
     with the three-parameter model given in Equation~\eqref{eq_F}, where
     a Gaussian damping envelope is distinguished from an exponential one.
 The switch time from one envelope to the other is represented
     by the vertical dash-dotted lines.
 For validation purposes, the parameter $\delta/s_0$ rather than $\delta$ itself
     is fixed (see the appendix for details). 
The flattening $\xi = 1-b/a$ labelled here represents the deviation of a cross-section from a circular one. 
}
\label{fig_app_M}
\end{figure}

\begin{figure}
\centering
\gridline{\fig{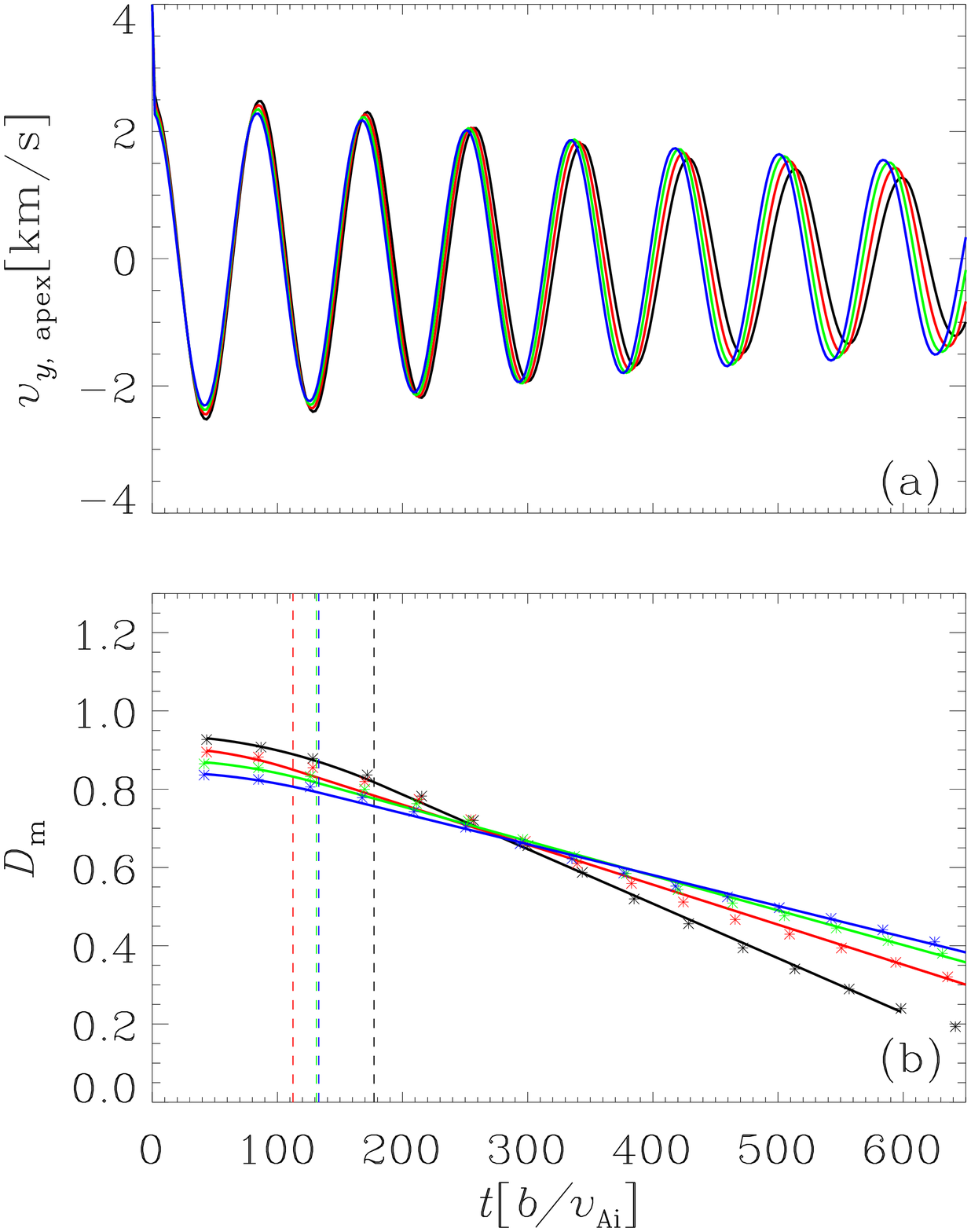}{0.9\textwidth}{}}
\caption{
  Similar to Figure \ref{fig_app_M}, but for kink oscillations polarized
     along the minor axis (m-modes).
  And hence the subscripts $y$ in $v_{y, {\rm apex}}$ and m in $D_{\rm m}$.   
}
\label{fig_app_m}
\end{figure}

\begin{figure}
\centering
\gridline{\fig{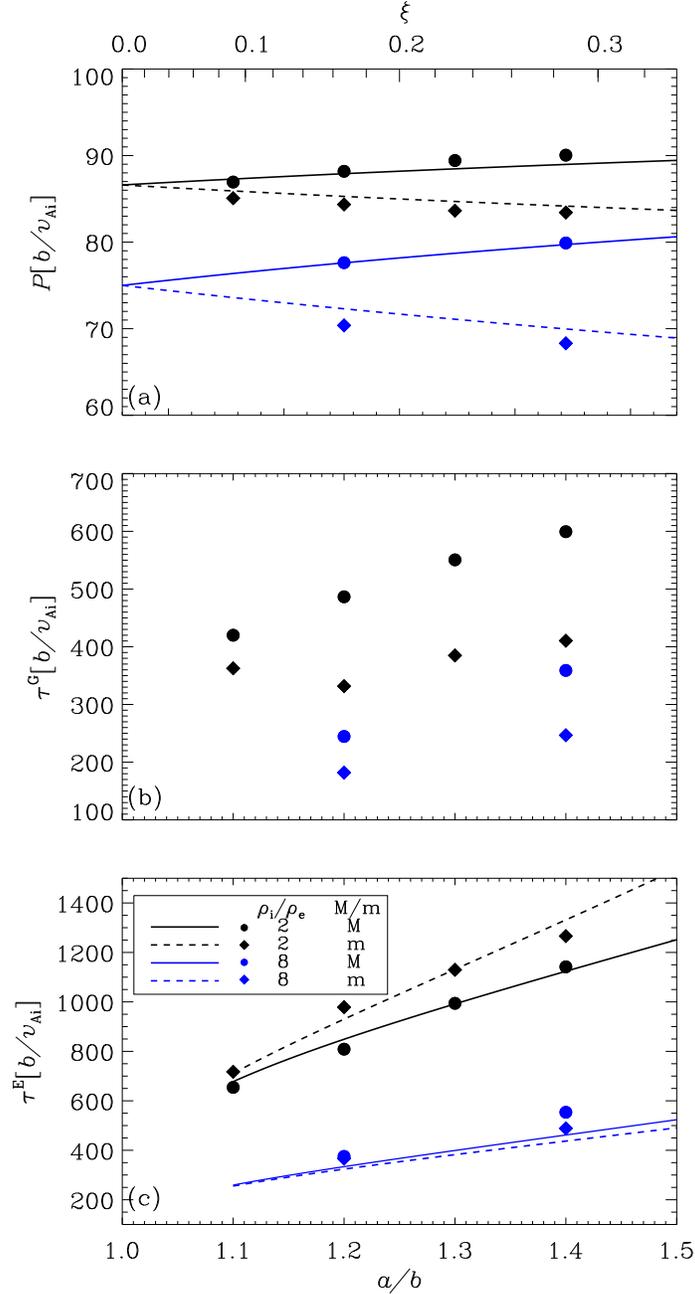}{0.5\textwidth}{}}
\caption{
  Dependence on the aspect ratio ($a/b$) of 
     (a) the periods ($P$)
     and (b, c) damping times ($\tau^{\rm G}, \tau^{\rm E}$)
     of the kink oscillations supported by loops with elliptic cross-sections and 
     linear transverse profiles described by Equation~\eqref{eq_app_fxy}. 
  Two different density ratios are examined as labeled by 
     the different colors. 
  The symbols represent the values found via best-fitting
     our numerical results.
  We use circles and diamonds to represent the M- and m-modes, respectively. 
  The curves represent the analytical expectations
     (Equations~\ref{eq_app_P} and \ref{eq_app_tau}) 
     from the eigen-mode analysis
     by~\citet{2003A&A...409..287R}, where the damping times pertain
     to the exponential stage.   
  We adopt the solid and dashed curves to represent the 
     M- and m-modes, respectively.         
  For validation purposes, the parameter
     $\delta/s_0$ rather than $\delta$ itself
     is fixed (see the appendix for details). 
Note that the flattening $\xi = 1-b/a$ labelled here represents the deviation of a cross-section from a circular one. 
}
\label{fig_app_R03}
\end{figure}

\begin{figure}
	\centering
	\gridline{\fig{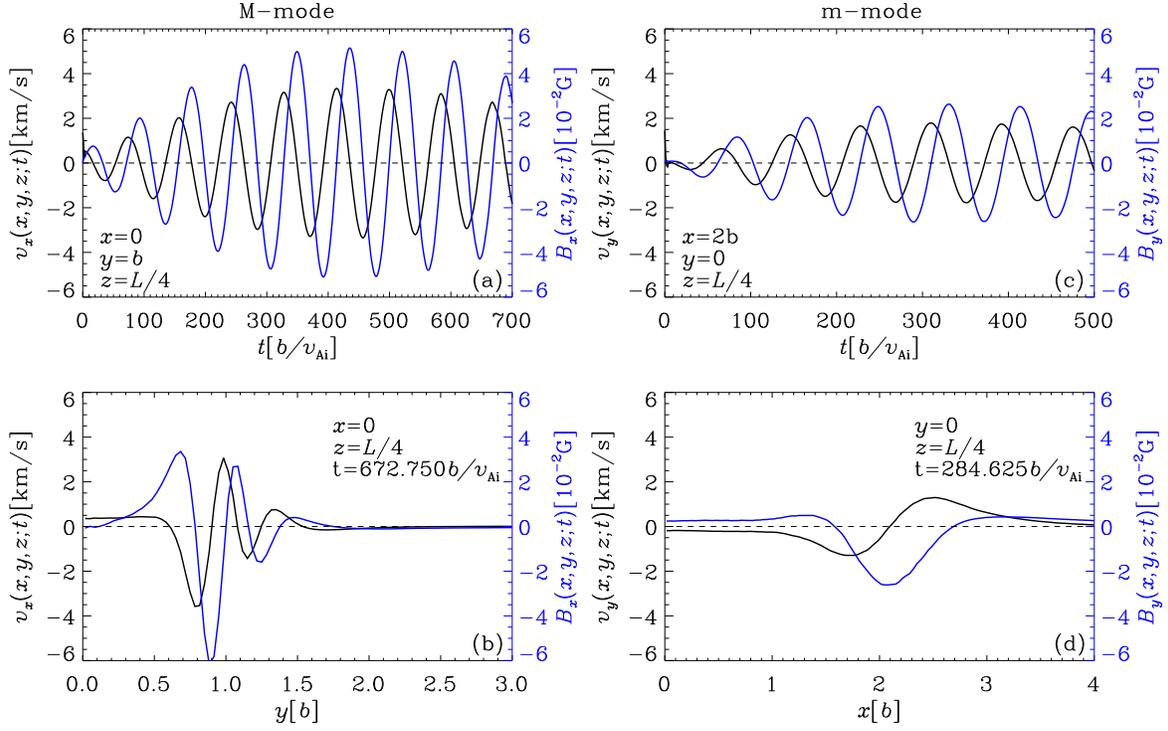}{0.85\textwidth}{}
	}
	\caption{
    Spatial and temporal distributions of the relevant
        component of the velocity (the black curves)
        and magnetic field vectors (blue), 
        sampled for examining
        the M-mode (the left column) and m-mode (right)
        kink oscillations in a loop with a density contrast of 
        $\rho_{\rm i}/\rho_{\rm e}=2$ and an aspect ratio
        of $a/b = 2$.
   Among the full set of
        the spatial and temporal coordinates,
        only one is allowed to vary in each panel,
        whereas the remaining ones are fixed as given by the numbers.
   This loop pertains to the equilibrium configuration examined in the main text.
	}			
	\label{fig_app2_alfven}
\end{figure}

\setcounter{equation}{0}
{
We have argued that the apparent damping of the kink oscillations is
    due to their resonant conversion into Alfv\'en modes
    in the inhomogeneous layer surrounding the loop boundary. 
While this argument is physically intuitive, 
    some further support seems desirable to make it more concrete. 
For this purpose we will choose a number of representative loops 
    that have been examined in the main text.
In addition, let us recall that the loops are primarily characterized
    by the density contrast $\rho_{\rm i}/\rho_{\rm e}$
    and the major-to-minor-half-axis-ratio $a/b$.

We start by shedding some light on the nature of
    the enhanced oscillations in the inhomgeneous layer
    pertaining to a loop with $\rho_{\rm i}/\rho_{\rm e} = 2$
    and $a/b=2$.
This loop is chosen because the corresponding M-mode (m-mode) has been detailed
    in Figures~\ref{fig_snapshots_Major} and \ref{fig_Major_v_cuts}
     (Figures~\ref{fig_snapshots_minor} and \ref{fig_minor_v_cuts}).
It suffices to examine the M-mode for now, for which the left column of
    Figure~\ref{fig_app2_alfven} presents some detailed behavior of
    the $x$-component of the velocity ($v_x$, the black curves)
    and magnetic field ($B_x$, blue).
Figure~\ref{fig_app2_alfven}a samples both quantities at a fixed location of
    $[x, y, z] = [0, b, L/4]$, while Figure~\ref{fig_app2_alfven}b provides
    a snapshot for both variables at $t = 672.75~b/v_{\rm Ai}$ sampled at
    $[x, z] = [0, L/4]$.
From Figure~\ref{fig_app2_alfven}a one sees that both $v_x$ and $B_x$ grow almost immediately
    after the impulsive leaky phase.
After a couple of cycles, $v_x$ tends to lead      
    $B_x$ by $\pi/2$.
In addition, Figure~\ref{fig_app2_alfven}a indicates that 
    a similar phase difference exists between 
    the $y$-distributions of
    the two quantities as well.
This phase difference is a strong indication of the Alfv\'enic nature of 
    the sampled oscillatory signals.
To see this, recall that only linear perturbations are of interest in the present study,
    and that our equilibrium magnetic field $\vec{B}_0$ is in the $z$-direction.
Recall further that the transverse structuring of 
    our equilibrium quantities is realized through $\bar{r}$
   (see Equation~\ref{eq_fr}).     
It then follows that, in the $x=0$ plane, 
   the $x$-direction is perpendicular to both $\vec{B}_0$ and
   the direction of inhomogeneity (the $y$-direction).
Linear Alfv\'en waves in this case are well-known to be governed by 
   \citep[e.g.,][]{1983A&A...117..220H} 
\begin{eqnarray}
\displaystyle 
  \frac{\partial v_{x}}{\partial t} 
& = & 
  \frac{B_0(y)}{\mu_0 \rho_0(y)}
  \frac{\partial B_{x}}{\partial z}~, 
  			\label{eq_app2_alfven1} \\
\displaystyle
  \frac{\partial B_{x}}{\partial t} 
& = &
  B_0(y)\frac{\partial v_{x}}{\partial z}~.
			\label{eq_app2_alfven2}
\end{eqnarray}
Given the boundary conditions, one readily finds that one solution
   to Equations~\eqref{eq_app2_alfven1} and \eqref{eq_app2_alfven2}
   reads
\begin{eqnarray}
&&   v_x(y, z; t) =  {\cal C}(y) v_{\rm A}(y) 
     \sin\left[k_z v_{\rm A}(y) t \right] 
     \sin\left(k_z z\right)~,
     \label{eq_Major_alfven_vx}\\
&&   B_x(y, z; t) = -{\cal C}(y) B_0(y)
     \cos\left[k_z v_{\rm A}(y) t \right] 
     \cos\left(k_z z\right)~,
     \label{eq_Major_alfven_Bx}
\end{eqnarray}  
   where $k_z = \pi/L$ is the axial wavenumber,
   and the $y$-dependence of the amplitude function 
   ${\cal C}(y)$ is in principle 
   arbitrary at this point.  
Evidently, this solution represents an axial fundamental mode,
   and therefore
   $v_x$ leads $B_x$ by $\pi/2$ for a given pair of $[y, z]$ when 
   $0 < z < L/2$. 
Our numerical results are consistent with this expectation.
In fact, the reason that we have chosen to sample the signals somewhere between 
    the footpoint ($z=0$) and apex ($z=L/2$) is to make sure
    that $v_x$ and $B_x$ are both finite for the ease of presentation.
Furthermore, it turns out that the spatial dependence of the Alfv\'en speed $v_{\rm A}(y)$ 
    (or rather $v_{\rm A}(\bar{r})$) 
    in the inhomgeneous layer 
    is not far from a linear one.
On top of that, the envelope 
   ${\cal C}(y)$ varies on a spatial scale larger than the layer width.
It then follows from Equations~\eqref{eq_Major_alfven_vx}   
   and \eqref{eq_Major_alfven_Bx} that
   $v_x(y)$ leads $B_x(y)$ by roughly $\pi/2$ 
   at some given $[x, z; t]$ when $0 < z < L/2$.
Note that the particular form
   in Equations~\eqref{eq_Major_alfven_vx}   
     and \eqref{eq_Major_alfven_Bx}
  is an indication that these linear Alfv\'en waves
  are connected to the kink oscillations.
After all, that they appear as an axial fundamental derives from
  the imposed $z$-dependence of our velocity driver.
We note further that the $y$-dependence of $v_{\rm A}$ is responsible for the phase-mixing
  of the resonantly converted Alfv\'en waves, an effect well-known since 
 \citeauthor{1983A&A...117..220H}~(\citeyear{1983A&A...117..220H}; see also \citeauthor{2015ApJ...812..121K}~\citeyear{2015ApJ...812..121K} for the consequences of
  phase-mixing).
However, moving away from the $x=0$ plane, 
  the Alfv\'enic nature of the enhanced oscillations cannot be revealed in a similar manner.
Rather, it can be told from the close resemblance of 
  the flow pattern presented in Figure~\ref{fig_snapshots_Major} to the $m=1$
  Alfv\'en waves \citep[][Figure~1]{1982SoPh...75....3S} in loops with circular cross-sections,
  despite that we have examined coronal loops
  with elliptic cross-sections and hence the azimuthal wavenumber $m$ cannot be
  unambiguously defined.
All the discussions we have offered apply to the m-mode as well, for which 
  the right column of 
  Figure~\ref{fig_app2_alfven} examines $v_y$ and $B_y$ that are
  sampled in the $y=0$ plane. 
}

\begin{figure}
	\centering
	\gridline{\fig{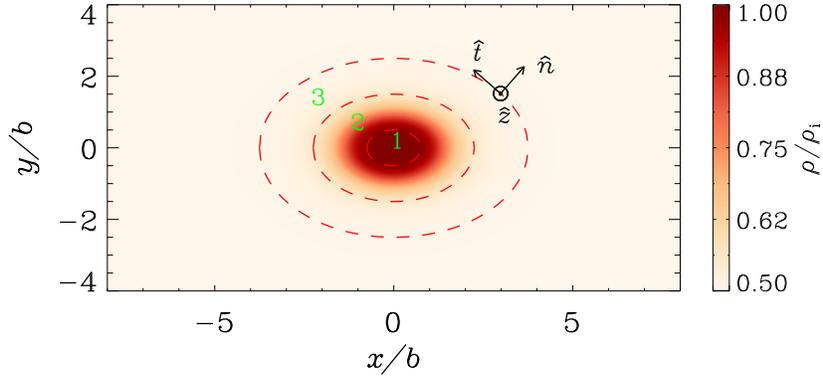}{0.6\textwidth}{}
	}
	\caption{
	Illustration of the coordinate system employed for examining the energetics
	    of the loop system. 
	Shown for reference is the $xy$-distribution of the equilibrium density 
	    for a loop with 
	    a density contrast of 
	        $\rho_{\rm i}/\rho_{\rm e}=2$ and an aspect ratio
	        of $a/b = 1.5$.    
    This loop pertains to the equilibrium configuration examined in the main text.
    The intersection of any magnetic surface with the $xy$-plane is an ellipsis,
        and can be labeled by the dimensionless variable $\bar{r}$
        (see Equation~\ref{eq_fr}).
    In the $xy$-plane, the unit vectors normal and tangential
        to a magnetic surface are denoted by $\hat{n}$ and $\hat{t}$, 
        respectively.    
    For later use, three regions are distinguished as enclosed by a selected number
        of magnetic surfaces shown by the dashed curves. 
    The outer boundaries of regions~1 to 3 correspond to 
        $\bar{r} = 0.5, 1.5$, and $2.5$, respectively.
    The inhomogeneity of the equilibrium parameters is contained in region~2 only.
	}			
	\label{fig_app2_density}
\end{figure}

{
The resonant conversion of kink oscillations to localized Alfv\'en modes
    can also be examined from the energetics perspective.
This examination turns out to be non-trivial.
We start by noting that 
    it is well established that
    the total energy density ($\epsilon$)
    and energy flux density ($\vec{f}$) in general do not adequately
    capture the wave energetics~\citep[e.g.,][]{2003ApJ...599..626B,2009A&A...508..951V}.
While some modified $\epsilon$ and $\vec{f}$ targeting wave-like
    perturbations
    have been developed 
    (\citeauthor{1994PhPl....1..691W}~\citeyear{1994PhPl....1..691W};
    see also 
    \citeauthor{1965RvPP....1..205B}~\citeyear{1965RvPP....1..205B}
    \citeauthor{1974soch.book.....B}~\citeyear{1974soch.book.....B}),     
    they suffer from interpretative ambiguities
    \citep[see Section~4 in][for a detailed discussion]{2003ApJ...599..626B}
    and need to be extended for the equilibrium configuration examined in 
    this study. 
We choose to start directly from 
    the linearized ideal MHD equations by maintaining mathematical consistency at this stage,
    and leave more insightful physical interpretations for a future study.   

A conservation law can be readily found for linear wave-like perturbations,
\begin{eqnarray}
\displaystyle
\frac{\partial \epsilon }{\partial t} = -\nabla\cdot\vec{f} + s~,
\label{eq_energy}
\end{eqnarray}    
   where $\epsilon$ and $\vec{f}$ represent some wave-related energy 
   and energy flux densities, respectively.
Defined by 
\begin{eqnarray}
\displaystyle 
 \epsilon 
= 
   \frac{1}{2}\rho_0\delta\vec{v}^2
 + \frac{\delta\vec{B}^2}{2\mu_0}
 + \frac{(\delta p)^2}{2\gamma p_0}~~,
  			\label{eq_epsilon} \\
\displaystyle 
  \vec{f} 
= 
  \delta p \delta \vec{v}
+ \frac{\delta\vec{B}\times \left(\delta\vec{v}\times\vec{B_0}\right)}{\mu_0}~,
  \label{eq_flux}
\end{eqnarray}
   they agree with the well-known expressions as offered in, say, \citet{2003ApJ...599..626B}.
Here the symbols with a $\delta$ represent the small-amplitude perturbations
   to the equilibrium, and can be explicitly written as
   $\delta p(\vec{x}; t) = p(\vec{x}; t) - p_0(\vec{x})$, 
   $\delta \vec{v}(\vec{x}; t) = \vec{v}(\vec{x}; t)$,
   and
   $\delta \vec{B}(\vec{x}; t) = \vec{B}(\vec{x}; t) - \vec{B}_0(\vec{x})$.
Different from the well-known results is 
  some source term
\begin{eqnarray}
s = \frac{1}{\mu_0}\left[(\nabla\times\vec{B_0})\times\delta \vec{B}\right]
\cdot\delta\vec{v}-\frac{\delta p\delta\vec{v}\cdot{\nabla\ln p_0}}{\gamma}~.
\label{eq_source}
\end{eqnarray}
Evidently, $s$ arises only for nonuniform equilibria.
To be more precise, it is relevant only in those portions of an equilibrium 
   where the equilibrium quantities are not uniform.
From the energetics perspective, 
   it also makes sense to integrate Equation~\eqref{eq_energy} over some volume $V$,
   the result being
\begin{eqnarray}
\displaystyle 
   \frac{\rm d}{{\rm d}t}
    \int_V \epsilon {\rm d} V
 = -\oint_{\partial V} {\rm d}\vec{A} \cdot \vec{f} 
   +\int_V s {\rm d}V~,
   \label{eq_app2_int_cons_law}
\end{eqnarray}    
   where $\partial V$ is the surface that encloses $V$.
In addition, the surface integral collects the net flux that leaves
   the volume.  
   
Given our equilibrium configuration, one naturally chooses a volume $V$ 
   to be a straight cylinder that is concentric with the loop and is
   bounded by the planes $z=0$ and $z=L$.    
With our boundary conditions at the bounding planes, 
   it can be readily shown that 
   $f_z$ vanishes therein and hence the net flux in Equation~\eqref{eq_app2_int_cons_law}
   derives only from the lateral surface of 
   the volume $V$.
Physically speaking, 
   this lateral surface is a magnetic surface, namely where
   the Alfv\'en speed $v_{\rm A}$ or equivalently
   $\bar{r}$ is constant (see Equation~\ref{eq_r}).
For mathematical convenience, 
   we then
   introduce a new coordinate system $(\bar{r}, \phi)$
   in the $xy$-plane such that the 
   Cartesian coordinates $(x, y)$ of any point 
   can be expressed by
\begin{eqnarray}
   x = a \bar{r}\cos\phi~, 
   y = b \bar{r}\sin\phi~,
\label{eq_coor_magsurface}
\end{eqnarray}    
   where $\phi$ ranges between $0$ and $2\pi$.
Note that this $\phi$ does not have the same meaning as in
   a standard cylindrical coordinate system. 
Regardless, the unit vector $\hat{n}$ normal to a magnetic surface
   can be readily evaluated as
\begin{eqnarray}
   \hat{n} = \frac{b\cos\phi \hat{x} +  
    a\sin\phi \hat{y}}{\sqrt{a^2\sin^2\phi+b^2\cos^2\phi}}~.
   \label{eq_app2_vector}
\end{eqnarray}
This unit vector is shown in Figure~\ref{fig_app2_density}, 
   together with the unit vector tangential
   to a magnetic surface ($\hat{t}$)
   and the spatial distribution
   of the equilibrium density $\rho_0$ in the $xy$-plane.
Note that the ellipses in Figure~\ref{fig_app2_density} correspond to
   a selected number of magnetic surfaces.    
To evaluate the integrals in Equation~\eqref{eq_app2_int_cons_law},
   we note further that the line elements in the $\hat{t}$- and 
   $\hat{n}$-directions (${\rm d}l_t$ and ${\rm d}l_n$) read
\begin{eqnarray}
\displaystyle 
&& {\rm d}l_t = \sqrt{a^2\sin^2\phi+b^2\cos^2\phi}  (\bar{r} \mathd \phi)~, \\
\displaystyle
&& {\rm d}l_n = \frac{ab}{\sqrt{a^2\sin^2\phi+b^2\cos^2\phi}} \mathd \bar{r}~.
\end{eqnarray} 
Now labeling the lateral surface of the volume with $\bar{r}$, 
   one readily recognizes that 
\begin{eqnarray}
  \int_V g(x, y, z; t) \mathd V 
= \int_0^{L} \mathd z 
  \int_0^{\bar{r}} (ab) \bar{r}' \mathd \bar{r}'
  \int_0^{2\pi} g(\bar{r}', \phi, z; t) \mathd \phi~,    
\end{eqnarray}   
   where $g$ represents both $\epsilon$ and $s$.
Likewise, one readily sees that 
\begin{eqnarray}
  \oint_{\partial V} {\rm d}\vec{A} \cdot \vec{f}(x, y, z; t)
= \int_0^{L} \mathd z 
  \int_0^{2\pi} \hat{n} \cdot \vec{f}(\bar{r}, \phi, z; t) 
  \sqrt{a^2\sin^2\phi+b^2\cos^2\phi}  (\bar{r} \mathd \phi)~.    
\end{eqnarray}   
With the following definitions
\begin{eqnarray}
&& E(\bar{r}; t) = \int_V \epsilon \mathd V~,~~
   S(\bar{r}; t) = \int_V s \mathd V~,~~  
   \Sigma(\bar{r}; t) = \int_{0}^{t} dt' S(\bar{r}; t')~, \nonumber \\[-0.2cm]
&& \label{eq_app2_cons_defs_intgral}    \\[-0.2cm]
&& F(\bar{r}; t) = \oint_{\partial V} {\rm d}\vec{A} \cdot \vec{f}~,~~~
   \Phi(\bar{r}; t) = \int_{0}^{t} dt' F(\bar{r}; t')~, \nonumber 
\end{eqnarray}
   it then follows from Equation~\eqref{eq_app2_int_cons_law}
   that 
\begin{eqnarray}
\displaystyle 
\frac{\mathd E(\bar{r}; t)}{\mathd t} = -F(\bar{r}; t) + S(\bar{r}; t)~,
\label{eq_app2_cons_dEdt}
\end{eqnarray}   
   or equivalently
\begin{eqnarray}
\displaystyle 
E(\bar{r}; t) = E(\bar{r}; t=0)-\Phi(\bar{r}; t) + \Sigma(\bar{r}; t)~.
\label{eq_app2_cons_EF}
\end{eqnarray}   
   
To proceed, we rather arbitrarily
   choose a loop with $\rho_{\rm i}/\rho_{\rm e}=2$ and $a/b = 1.5$,
   and examine the M-mode only. 
Shown in Figure~\ref{fig_app2_density}, three regions in the plane transverse to this
   loop are distinguished as labeled.
Let $\bar{r}_j$ denote the outer boundary of each region ($j=1, 2, 3$).
For the ease of description, the following shorthand notations prove helpful,
\begin{eqnarray}
   F_j    = F(\bar{r}_j; t)~,~~~
   \Phi_j = \Phi(\bar{r}_j; t)~,
\end{eqnarray}
   which pertain to the energy flux leaving the outer boundary of region $j$.
Likewise, we adopt the shorthand notions $E_j$ and $\Sigma_j$ to denote
   the contribution to the relevant volume-integrals from only region $j$.
Evidently,    
\begin{eqnarray}
   E_j       = E(\bar{r}_{j}; t) - E(\bar{r}_{j-1}; t),~~~
   \Sigma_j  = \Sigma(\bar{r}_{j}; t) - \Sigma(\bar{r}_{j-1}; t)~,
\end{eqnarray}
   when $j=2, 3$.
Furthermore, $E_1 = E(\bar{r}_1; t)$ and 
   $\Sigma_1  = \Sigma(\bar{r}_1; t)$.
We choose $\bar{r}_1 = 0.5$ and $\bar{r}_2 = 1.5$ such that the inhomogeneity
   is essentially present only in region~2 (recall that the nominal loop boundary
   corresponds to $\bar{r} = 1$). 
In practice, this means that $\Sigma_j \ne 0$ only when $j=2$.      
In addition, we set $\bar{r}_3 = 2.5$ for the following reason.
Drawing analogy with kink oscillations in loops with circular cross-sections, one expects
   that the apparent damping of the kink oscillations in our case is connected to 
   some net energy flow into region~2 
   \citep[see e.g.,][]{1994PhPl....1..691W, 2013ApJ...777..158S}.
However, for this to be evaluated more quantitatively, 
   one would require that $F_3$
   be negligible after the kink oscillations have been set up.

Figure~\ref{fig_app2_energyconsv} examines the temporal evolution of 
   the time-integrated energy flux ($\Phi_3$, the black dashed curve).
Note that for the ease of presentation, 
   here $E(\bar{r}_3; t=0)-\Phi(\bar{r}_3; t) = (E_1+E_2+E_3)|_{t=0}-\Phi_3$
   rather than $\Phi_3$ 
   is plotted.  
Regardless, the point is that this black dashed curve should 
   become close to a horizontal line provided that  
   $F_3$ vanishes after the system has
   evolved for some time
   (see Equation~\ref{eq_app2_cons_defs_intgral}).
In fact, this expectation takes place almost immediately after the impulsive leaky phase.
On top of that, this behavior is true even if we choose a larger value of $\bar{r}_3$,
   which is nonetheless rather arbitrarily set to $2.5$.
Figure~\ref{fig_app2_energyconsv} also shows the temporal evolution
   of $E(\bar{r}_3; t)-\Sigma(\bar{r}_3; t)$ (the black solid curve),
   which is labeled with a subscript 1+2+3 because this quantity
   incorporates the contributions from all the three regions. 
In view of Equation~\eqref{eq_app2_cons_EF}, one expects that
   the black solid and dashed curves should coincide with each other.
This is indeed true for $t \lesssim 280~b/v_{\rm Ai}$. 
In particular, the rapid initial drop in both curves offers a more insightful
   illustration of the impulsive leaky phase, namely the total energy in the volume
   bordered by $\bar{r}_3$ is lost through the outward energy flux.
One may then question whether the disagreement between the black solid and dashed curves
   means that our numerical results are not that trustworthy  
   for large $t$.
We address this issue by performing another numerical run with a substantially finer grid, 
   with the smallest grid size reaching $20$~km ($600$~km) in the $x$- and $y$-directions
   (the $z$-direction).
The corresponding results, plotted by the blue curves, indeed indicate a better agreement
   between the solid and dashed curves.
A relative difference of $\lesssim 25\%$ is seen, which is remarkably good
   for fully three-dimensional
   simulations. 
In contrast, the relative difference may reach $\sim 50\%$ for the coarse grid, namely
   the reference grid setup that has been consistently used in the main text.    

Now a digression from the energetics is necessary, because one may ask
   whether the periods and damping times are substantially different for different grid setups.
Figure~\ref{fig_app2_DM} examines the temporal evolution of $v_x$ sampled at the loop apex
   ($[x, y, z]=[0, 0, L/2]$) by showing both (a) the $v_{x, {\rm apex}}$ series     
   and (b) its damping envelope ($D_{\rm M}$).
Comparing the coarse-grid (the black curves) results with the 
   fine-grid ones (blue), one discerns some difference 
   only in $D_{\rm M}$ when $t\gtrsim 400~b/v_{\rm Ai}$.
We then perform the three-parameter fitting to the fine-grid version of $D_{\rm M}$,
   finding that the period ($P$)
   and the damping time in the Gaussian stage ($\tau^{\rm G}$)
   differ little from the coarse-grid values.
In fact, the damping time in the exponential stage ($\tau^{\rm E}$)
   is not that different either.
The relative difference is merely $9.8\%$, 
   which is much smaller than the differences in the relative accuracy
   regarding the overall energy balance.               
This means that it is much less numerically demanding for one to capture the 
   characteristic timescales for kink oscillations than to realize a remarkably
   accurate energy conservation.    
From this we conclude that the values we derived for $\tau^{\rm E}$
   with the reference grid setup are in general trustworthy.
On top of that, the derived values for $P$ and $\tau^{\rm G}$
   are even more reliable. 
That said, we note by passing that the fine-grid computation
   suggests that 
   the irregular oscillations
   toward the end of the simulations are likely to be primarily of numerical origin.
On the one hand, the exponential stage in this case does extend
   to some longer time than in the coarse-grid case.
On the other hand, the accuracy in energy conservation is less good
   than in the interval where the oscillations are regular.           

Now we are ready to examine the energy injection into region~2.
For this purpose we employ the mathematical preparation
    in Equation~\eqref{eq_app2_cons_dEdt} rather than Equation~\eqref{eq_app2_cons_EF},
    because the flux $F$ proves more intuitive than its time-integrated counterpart $\Phi$.
Figure~\ref{fig_app2_energy13} examines both (a) the total energy ($E_1 + E_3$)
    in regions~1 and 3
    and (b) the energy flux ($F_1-F_2$) injected into region~2 .
The reason for us to examine the energy in regions~1 and 2 is that
    the source term $S$ vanishes therein.     
As has been pointed out, the definition of $E$, which derives from the definition
    of the energy density $\epsilon$ (Equation~\ref{eq_epsilon}), 
    does not suffer any interpretive ambiguity.
However, while mathematically consistent, 
    the source term needs to be interpreted in a more physical manner.          
Leaving this aspect to a future dedicated study, we note that 
    the rate of decrease (increase) of $E_1 + E_2$
    is expected to be balanced by a positive (negative) $F_1-F_2$ 
    after the impulsive leaky phase.
Figure~\ref{fig_app2_energy13} indicates that     
    this expectation is indeed valid.
In particular, the overall loss of the wave energy in regions~1 and 2
    is due to a net positive $(F_1-F_2)$, thereby making it more concrete to
    conclude that the energy associated with the kink oscillation
    is resonantly absorbed somewhere in the inhomogeneous layer.
We note that the oscillatory behavior at half the kink period, which is particularly pronounced 
    in Figure~\ref{fig_app2_energy13}b, arises because we are evaluating
    second-order quantities (see Equations~\ref{eq_epsilon} and \ref{eq_flux}).
In addition, the fine-grid results (the blue curves)
    are inconsistent with the coarse-grid ones only when $t$ exceeds, say, 
    $\sim 400~b/v_{\rm Ai}$.
We have argued that this does not mean that the characteristic timescales
    we have derived with the reference grid setup is not trustworthy.
}

\clearpage 

\clearpage 

\clearpage 

\clearpage 

\clearpage 

\clearpage 

\clearpage 

\clearpage 

\clearpage 

\clearpage 

\clearpage 

\clearpage 

\clearpage 

\clearpage 
\begin{figure}
	\centering
	\gridline{\fig{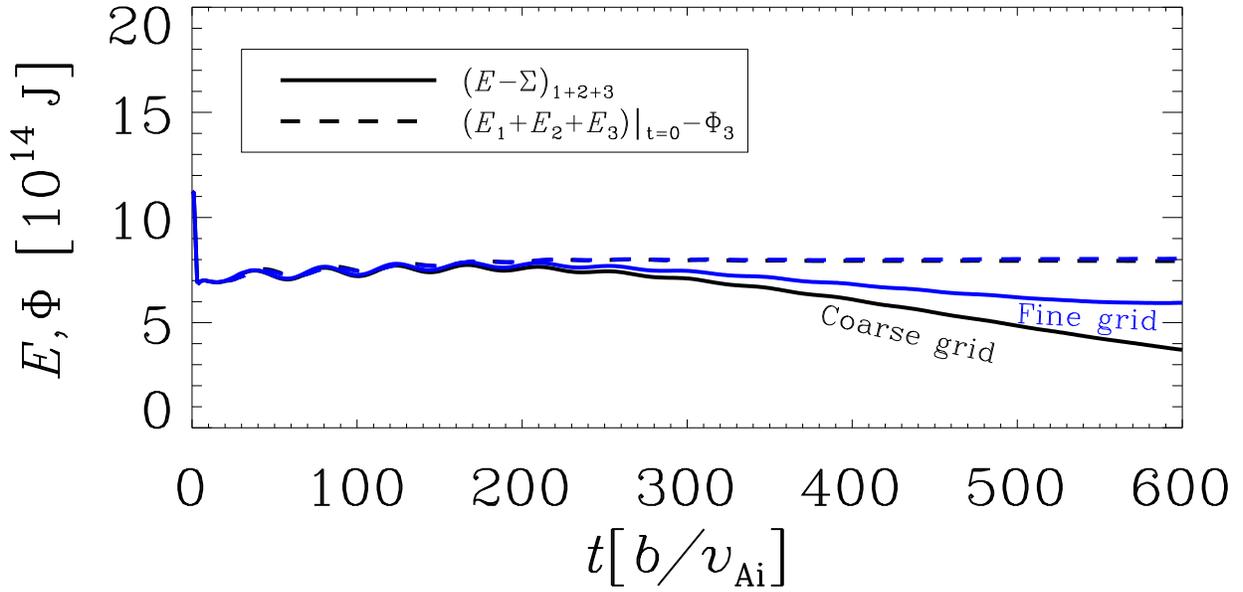}{0.9\textwidth}{}
	}
	\caption{
	Overall energy balance associated with wave-like perturbations in 
	    all the three regions in Figure~\ref{fig_app2_density}.
	Two computations are examined, one with the reference grid setup adopted
	    in the main text (the black curves), the other
	    with a substantially finer grid (blue).
	The solid curves represent $E-\Sigma$, which 
		characterizes the total energy content.
	The dashed curves are connected to the outward energy
	    flux leaving the outer boundary of region~3.
	Here the M-mode oscillation is examined, see Appendix~\ref{sec_appendix2}
	    for details.
	}			
	\label{fig_app2_energyconsv}
\end{figure}

\clearpage 
\begin{figure}
	\centering
	\gridline{\fig{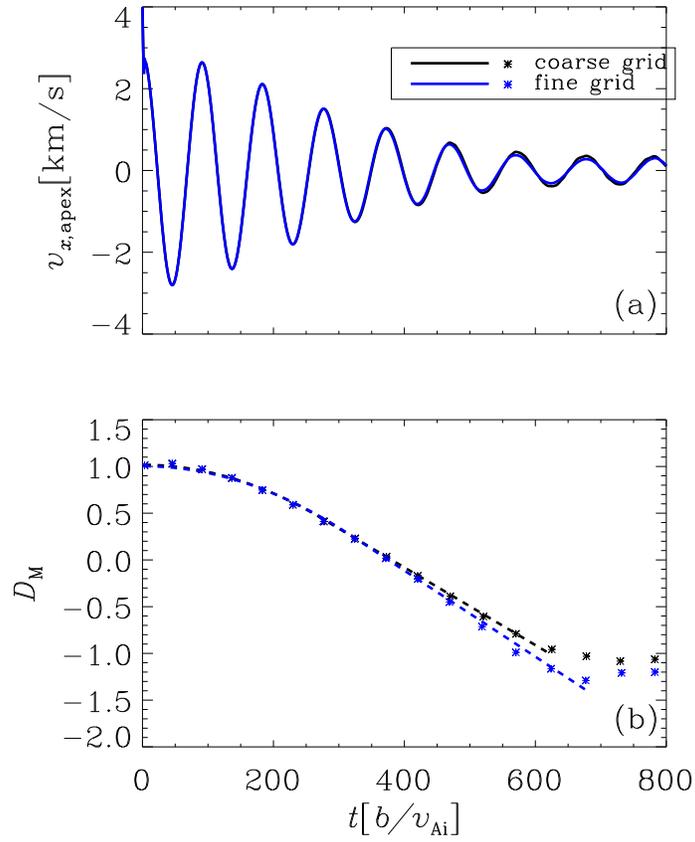}{0.5\textwidth}{}
	}
	\caption{
		 Similar to Figure \ref{fig_vx_t}, 
		 but for an M-mode kink oscillation in a loop
          with $a/b=1.5$ and $\rho_{\rm i}/\rho_{\rm i}=2$.
         The black curves here are reproduced from the curves
             pertinent to $a/b=1.5$ shown in Figure~\ref{fig_vx_t},
             while the blue lines 
             represent a fine-grid simulation.
	}			
	\label{fig_app2_DM}
\end{figure}

\clearpage 
\begin{figure}
	\centering
	\gridline{\fig{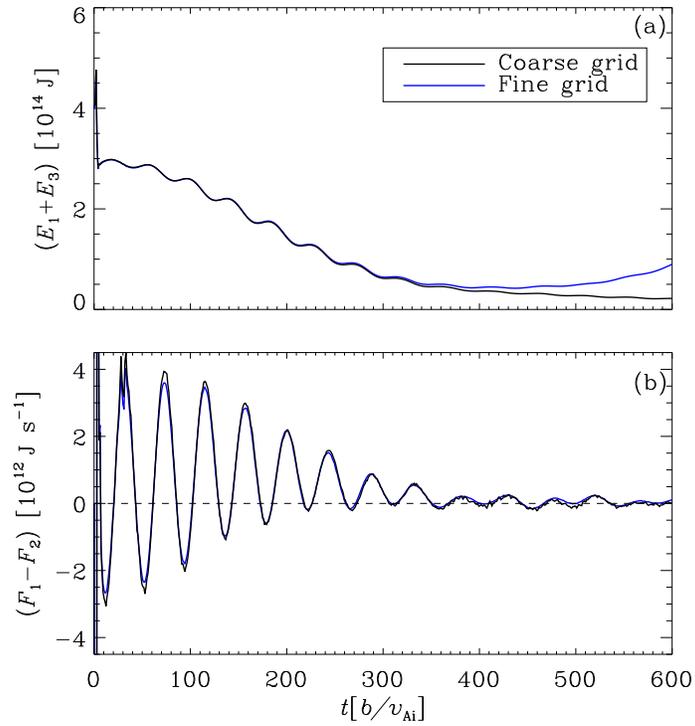}{0.5\textwidth}{}
	}
	\caption{
	Energy balance associated with wave-like perturbations in 
	    the three regions in Figure~\ref{fig_app2_density}.
	Shown here are the temporal profiles of (a) the total energy in regions~1 and 3,
	    and (b) the net energy flux injected into region~2. 
	Here the M-mode oscillation is examined, see Appendix~\ref{sec_appendix2}
	    for details.
	}			
	\label{fig_app2_energy13}
\end{figure}

\clearpage
\bibliographystyle{aasjournal}
\bibliography{ellip_loop_v2.0}

\begin{thebibliography}{}
\expandafter\ifx\csname natexlab\endcsname\relax\def\natexlab#1{#1}\fi
\providecommand{\url}[1]{\href{#1}{#1}}

\bibitem[{{Afanasyev} {et~al.}(2020){Afanasyev}, {Van Doorsselaere}, \&
  {Nakariakov}}]{2020A&A...633L...8A}
{Afanasyev}, A.~N., {Van Doorsselaere}, T., \& {Nakariakov}, V.~M. 2020, \aap,
  633, L8

\bibitem[{{Andries} {et~al.}(2005){Andries}, {Arregui}, \&
  {Goossens}}]{2005ApJ...624L..57A}
{Andries}, J., {Arregui}, I., \& {Goossens}, M. 2005, \apjl, 624, L57

\bibitem[{{Andries} {et~al.}(2009){Andries}, {Van Doorsselaere}, {Roberts},
  {Verth}, {Verwichte}, \& {Erd{\'e}lyi}}]{2009SSRv..149....3A}
{Andries}, J., {Van Doorsselaere}, T., {Roberts}, B., {et~al.} 2009, \ssr, 149,
  3

\bibitem[{{Anfinogentov} {et~al.}(2013){Anfinogentov}, {Nistic{\`o}}, \&
  {Nakariakov}}]{2013A&A...560A.107A}
{Anfinogentov}, S., {Nistic{\`o}}, G., \& {Nakariakov}, V.~M. 2013, \aap, 560,
  A107

\bibitem[{{Anfinogentov} \& {Nakariakov}(2019)}]{2019ApJ...884L..40A}
{Anfinogentov}, S.~A., \& {Nakariakov}, V.~M. 2019, \apjl, 884, L40

\bibitem[{{Anfinogentov} {et~al.}(2015){Anfinogentov}, {Nakariakov}, \&
  {Nistic{\`o}}}]{2015A&A...583A.136A}
{Anfinogentov}, S.~A., {Nakariakov}, V.~M., \& {Nistic{\`o}}, G. 2015, \aap,
  583, A136

\bibitem[{{Antolin} {et~al.}(2016){Antolin}, {De Moortel}, {Van Doorsselaere},
  \& {Yokoyama}}]{2016ApJ...830L..22A}
{Antolin}, P., {De Moortel}, I., {Van Doorsselaere}, T., \& {Yokoyama}, T.
  2016, \apjl, 830, L22

\bibitem[{{Antolin} {et~al.}(2015){Antolin}, {Okamoto}, {De Pontieu},
  {Uitenbroek}, {Van Doorsselaere}, \& {Yokoyama}}]{2015ApJ...809...72A}
{Antolin}, P., {Okamoto}, T.~J., {De Pontieu}, B., {et~al.} 2015, \apj, 809, 72

\bibitem[{{Arregui} {et~al.}(2007){Arregui}, {Andries}, {Van Doorsselaere},
  {Goossens}, \& {Poedts}}]{2007A&A...463..333A}
{Arregui}, I., {Andries}, J., {Van Doorsselaere}, T., {Goossens}, M., \&
  {Poedts}, S. 2007, \aap, 463, 333

\bibitem[{{Arregui} \& {Goossens}(2019)}]{2019A&A...622A..44A}
{Arregui}, I., \& {Goossens}, M. 2019, \aap, 622, A44

\bibitem[{{Arregui} {et~al.}(2011){Arregui}, {Soler}, {Ballester}, \&
  {Wright}}]{2011A&A...533A..60A}
{Arregui}, I., {Soler}, R., {Ballester}, J.~L., \& {Wright}, A.~N. 2011, \aap,
  533, A60

\bibitem[{{Aschwanden}(2009)}]{2009SSRv..149...31A}
{Aschwanden}, M.~J. 2009, \ssr, 149, 31

\bibitem[{{Aschwanden}(2011)}]{2011LRSP....8....5A}
---. 2011, Living Reviews in Solar Physics, 8, 5

\bibitem[{{Aschwanden} {et~al.}(1999){Aschwanden}, {Fletcher}, {Schrijver}, \&
  {Alexander}}]{1999ApJ...520..880A}
{Aschwanden}, M.~J., {Fletcher}, L., {Schrijver}, C.~J., \& {Alexander}, D.
  1999, \apj, 520, 880

\bibitem[{{Aschwanden} {et~al.}(2004){Aschwanden}, {Nakariakov}, \&
  {Melnikov}}]{2004ApJ...600..458A}
{Aschwanden}, M.~J., {Nakariakov}, V.~M., \& {Melnikov}, V.~F. 2004, \apj, 600,
  458

\bibitem[{{Aschwanden} {et~al.}(2003){Aschwanden}, {Nightingale}, {Andries},
  {Goossens}, \& {Van Doorsselaere}}]{2003ApJ...598.1375A}
{Aschwanden}, M.~J., {Nightingale}, R.~W., {Andries}, J., {Goossens}, M., \&
  {Van Doorsselaere}, T. 2003, \apj, 598, 1375

\bibitem[{{Aschwanden} \& {Schrijver}(2011)}]{2011ApJ...736..102A}
{Aschwanden}, M.~J., \& {Schrijver}, C.~J. 2011, \apj, 736, 102

\bibitem[{{Banerjee} {et~al.}(2007){Banerjee}, {Erd{\'e}lyi}, {Oliver}, \&
  {O'Shea}}]{2007SoPh..246....3B}
{Banerjee}, D., {Erd{\'e}lyi}, R., {Oliver}, R., \& {O'Shea}, E. 2007,
  \solphys, 246, 3

\bibitem[{{Bogdan} {et~al.}(2003){Bogdan}, {Carlsson}, {Hansteen}, {McMurry},
  {Rosenthal}, {Johnson}, {Petty-Powell}, {Zita}, {Stein}, {McIntosh}, \&
  {Nordlund}}]{2003ApJ...599..626B}
{Bogdan}, T.~J., {Carlsson}, M., {Hansteen}, V.~H., {et~al.} 2003, \apj, 599,
  626

\bibitem[{{Braginskii}(1965)}]{1965RvPP....1..205B}
{Braginskii}, S.~I. 1965, Reviews of Plasma Physics, 1, 205

\bibitem[{{Bray} \& {Loughhead}(1974)}]{1974soch.book.....B}
{Bray}, R.~J., \& {Loughhead}, R.~E. 1974, {The solar chromosphere}

\bibitem[{{Browning} \& {Priest}(1984)}]{1984A&A...131..283B}
{Browning}, P.~K., \& {Priest}, E.~R. 1984, \aap, 131, 283

\bibitem[{{Chen} {et~al.}(2015){Chen}, {Li}, {Xiong}, {Yu}, \&
  {Guo}}]{2015ApJ...812...22C}
{Chen}, S.-X., {Li}, B., {Xiong}, M., {Yu}, H., \& {Guo}, M.-Z. 2015, \apj,
  812, 22

\bibitem[{{De Moortel} \& {Nakariakov}(2012)}]{2012RSPTA.370.3193D}
{De Moortel}, I., \& {Nakariakov}, V.~M. 2012, Philosophical Transactions of
  the Royal Society of London Series A, 370, 3193

\bibitem[{{De Moortel} \& {Pascoe}(2009)}]{2009ApJ...699L..72D}
{De Moortel}, I., \& {Pascoe}, D.~J. 2009, \apjl, 699, L72

\bibitem[{{Duckenfield} {et~al.}(2019){Duckenfield}, {Goddard}, {Pascoe}, \&
  {Nakariakov}}]{2019A&A...632A..64D}
{Duckenfield}, T.~J., {Goddard}, C.~R., {Pascoe}, D.~J., \& {Nakariakov}, V.~M.
  2019, \aap, 632, A64

\bibitem[{{Dymova} \& {Ruderman}(2006)}]{2006A&A...457.1059D}
{Dymova}, M.~V., \& {Ruderman}, M.~S. 2006, \aap, 457, 1059

\bibitem[{{Edwin} \& {Roberts}(1983)}]{1983SoPh...88..179E}
{Edwin}, P.~M., \& {Roberts}, B. 1983, \solphys, 88, 179

\bibitem[{{Erd{\'e}lyi} \& {Morton}(2009)}]{2009A&A...494..295E}
{Erd{\'e}lyi}, R., \& {Morton}, R.~J. 2009, \aap, 494, 295

\bibitem[{{Erd{\'e}lyi} \& {Taroyan}(2008)}]{2008A&A...489L..49E}
{Erd{\'e}lyi}, R., \& {Taroyan}, Y. 2008, \aap, 489, L49

\bibitem[{{Goddard} {et~al.}(2016){Goddard}, {Nistic{\`o}}, {Nakariakov}, \&
  {Zimovets}}]{2016A&A...585A.137G}
{Goddard}, C.~R., {Nistic{\`o}}, G., {Nakariakov}, V.~M., \& {Zimovets}, I.~V.
  2016, \aap, 585, A137

\bibitem[{{Goossens} {et~al.}(2002){Goossens}, {Andries}, \&
  {Aschwanden}}]{2002A&A...394L..39G}
{Goossens}, M., {Andries}, J., \& {Aschwanden}, M.~J. 2002, \aap, 394, L39

\bibitem[{{Goossens} {et~al.}(2008){Goossens}, {Arregui}, {Ballester}, \&
  {Wang}}]{2008A&A...484..851G}
{Goossens}, M., {Arregui}, I., {Ballester}, J.~L., \& {Wang}, T.~J. 2008, \aap,
  484, 851

\bibitem[{{Goossens} {et~al.}(2011){Goossens}, {Erd{\'e}lyi}, \&
  {Ruderman}}]{2011SSRv..158..289G}
{Goossens}, M., {Erd{\'e}lyi}, R., \& {Ruderman}, M.~S. 2011, \ssr, 158, 289

\bibitem[{{Goossens} {et~al.}(2014){Goossens}, {Soler}, {Terradas}, {Van
  Doorsselaere}, \& {Verth}}]{2014ApJ...788....9G}
{Goossens}, M., {Soler}, R., {Terradas}, J., {Van Doorsselaere}, T., \&
  {Verth}, G. 2014, \apj, 788, 9

\bibitem[{{Goossens} {et~al.}(2013){Goossens}, {Van Doorsselaere}, {Soler}, \&
  {Verth}}]{2013ApJ...768..191G}
{Goossens}, M., {Van Doorsselaere}, T., {Soler}, R., \& {Verth}, G. 2013, \apj,
  768, 191

\bibitem[{{Guo} {et~al.}(2019{\natexlab{a}}){Guo}, {Van Doorsselaere},
  {Karampelas}, \& {Li}}]{2019ApJ...883...20G}
{Guo}, M., {Van Doorsselaere}, T., {Karampelas}, K., \& {Li}, B.
  2019{\natexlab{a}}, \apj, 883, 20

\bibitem[{{Guo} {et~al.}(2019{\natexlab{b}}){Guo}, {Van Doorsselaere},
  {Karampelas}, {Li}, {Antolin}, \& {De Moortel}}]{2019ApJ...870...55G}
{Guo}, M., {Van Doorsselaere}, T., {Karampelas}, K., {et~al.}
  2019{\natexlab{b}}, \apj, 870, 55

\bibitem[{{Guo} {et~al.}(2015){Guo}, {Erd{\'e}lyi}, {Srivastava}, {Hao},
  {Cheng}, {Chen}, {Ding}, \& {Dwivedi}}]{2015ApJ...799..151G}
{Guo}, Y., {Erd{\'e}lyi}, R., {Srivastava}, A.~K., {et~al.} 2015, \apj, 799,
  151

\bibitem[{{Heyvaerts} \& {Priest}(1983)}]{1983A&A...117..220H}
{Heyvaerts}, J., \& {Priest}, E.~R. 1983, \aap, 117, 220

\bibitem[{{Hillier} {et~al.}(2019){Hillier}, {Barker}, {Arregui}, \&
  {Latter}}]{2019MNRAS.482.1143H}
{Hillier}, A., {Barker}, A., {Arregui}, I., \& {Latter}, H. 2019, \mnras, 482,
  1143

\bibitem[{{Hollweg} \& {Yang}(1988)}]{1988JGR....93.5423H}
{Hollweg}, J.~V., \& {Yang}, G. 1988, \jgr, 93, 5423

\bibitem[{{Hood} {et~al.}(2013){Hood}, {Ruderman}, {Pascoe}, {De Moortel},
  {Terradas}, \& {Wright}}]{2013A&A...551A..39H}
{Hood}, A.~W., {Ruderman}, M., {Pascoe}, D.~J., {et~al.} 2013, \aap, 551, A39

\bibitem[{{Howson} {et~al.}(2019){Howson}, {De Moortel}, {Antolin}, {Van
  Doorsselaere}, \& {Wright}}]{2019A&A...631A.105H}
{Howson}, T.~A., {De Moortel}, I., {Antolin}, P., {Van Doorsselaere}, T., \&
  {Wright}, A.~N. 2019, \aap, 631, A105

\bibitem[{{Ionson}(1978)}]{1978ApJ...226..650I}
{Ionson}, J.~A. 1978, \apj, 226, 650

\bibitem[Kaneko et al.(2015)]{2015ApJ...812..121K} Kaneko, T., Goossens, M., Soler, R., et al.\ 2015, \apj, 812, 121

\bibitem[{{Karampelas} {et~al.}(2017){Karampelas}, {Van Doorsselaere}, \&
  {Antolin}}]{2017A&A...604A.130K}
{Karampelas}, K., {Van Doorsselaere}, T., \& {Antolin}, P. 2017, \aap, 604,
  A130

\bibitem[{{Klimchuk} {et~al.}(2000){Klimchuk}, {Antiochos}, \&
  {Norton}}]{2000ApJ...542..504K}
{Klimchuk}, J.~A., {Antiochos}, S.~K., \& {Norton}, D. 2000, \apj, 542, 504

\bibitem[{{Kucera} {et~al.}(2019){Kucera}, {Young}, {Klimchuk}, \&
  {DeForest}}]{2019ApJ...885....7K}
{Kucera}, T.~A., {Young}, P.~R., {Klimchuk}, J.~A., \& {DeForest}, C.~E. 2019,
  \apj, 885, 7

\bibitem[{{L{\'o}pez Fuentes} {et~al.}(2006){L{\'o}pez Fuentes}, {Klimchuk}, \&
  {D{\'e}moulin}}]{2006ApJ...639..459L}
{L{\'o}pez Fuentes}, M.~C., {Klimchuk}, J.~A., \& {D{\'e}moulin}, P. 2006,
  \apj, 639, 459

\bibitem[{{Magyar} \& {Nakariakov}(2020)}]{2020ApJ...894L..23M}
{Magyar}, N., \& {Nakariakov}, V.~M. 2020, \apjl, 894, L23

\bibitem[{{Magyar} \& {Van Doorsselaere}(2016)}]{2016A&A...595A..81M}
{Magyar}, N., \& {Van Doorsselaere}, T. 2016, \aap, 595, A81

\bibitem[{{Malanushenko} \& {Schrijver}(2013)}]{2013ApJ...775..120M}
{Malanushenko}, A., \& {Schrijver}, C.~J. 2013, \apj, 775, 120

\bibitem[{{McLaughlin} \& {Ofman}(2008)}]{2008ApJ...682.1338M}
{McLaughlin}, J.~A., \& {Ofman}, L. 2008, \apj, 682, 1338

\bibitem[{{Mignone} {et~al.}(2007){Mignone}, {Bodo}, {Massaglia}, {Matsakos},
  {Tesileanu}, {Zanni}, \& {Ferrari}}]{2007ApJS..170..228M}
{Mignone}, A., {Bodo}, G., {Massaglia}, S., {et~al.} 2007, \apjs, 170, 228

\bibitem[{{Morton} \& {Ruderman}(2011)}]{2011A&A...527A..53M}
{Morton}, R.~J., \& {Ruderman}, M.~S. 2011, \aap, 527, A53

\bibitem[{{Nakariakov} {et~al.}(2016{\natexlab{a}}){Nakariakov},
  {Anfinogentov}, {Nistic{\`o}}, \& {Lee}}]{2016A&A...591L...5N}
{Nakariakov}, V.~M., {Anfinogentov}, S.~A., {Nistic{\`o}}, G., \& {Lee}, D.~H.
  2016{\natexlab{a}}, \aap, 591, L5

\bibitem[{{Nakariakov} \& {Ofman}(2001)}]{2001A&A...372L..53N}
{Nakariakov}, V.~M., \& {Ofman}, L. 2001, \aap, 372, L53

\bibitem[{{Nakariakov} {et~al.}(1999){Nakariakov}, {Ofman}, {Deluca},
  {Roberts}, \& {Davila}}]{1999Sci...285..862N}
{Nakariakov}, V.~M., {Ofman}, L., {Deluca}, E.~E., {Roberts}, B., \& {Davila},
  J.~M. 1999, Science, 285, 862

\bibitem[{{Nakariakov} \& {Verwichte}(2005)}]{2005LRSP....2....3N}
{Nakariakov}, V.~M., \& {Verwichte}, E. 2005, Living Reviews in Solar Physics,
  2, 3

\bibitem[{{Nakariakov} {et~al.}(2016{\natexlab{b}}){Nakariakov}, {Pilipenko},
  {Heilig}, {Jel{\'\i}nek}, {Karlick{\'y}}, {Klimushkin}, {Kolotkov}, {Lee},
  {Nistic{\`o}}, {Van Doorsselaere}, {Verth}, \&
  {Zimovets}}]{2016SSRv..200...75N}
{Nakariakov}, V.~M., {Pilipenko}, V., {Heilig}, B., {et~al.}
  2016{\natexlab{b}}, \ssr, 200, 75

\bibitem[{{Nechaeva} {et~al.}(2019){Nechaeva}, {Zimovets}, {Nakariakov}, \&
  {Goddard}}]{2019ApJS..241...31N}
{Nechaeva}, A., {Zimovets}, I.~V., {Nakariakov}, V.~M., \& {Goddard}, C.~R.
  2019, \apjs, 241, 31

\bibitem[{{Nistic{\`o}} {et~al.}(2013){Nistic{\`o}}, {Nakariakov}, \&
  {Verwichte}}]{2013A&A...552A..57N}
{Nistic{\`o}}, G., {Nakariakov}, V.~M., \& {Verwichte}, E. 2013, \aap, 552, A57

\bibitem[{{Ofman} \& {Wang}(2008)}]{2008A&A...482L...9O}
{Ofman}, L., \& {Wang}, T.~J. 2008, \aap, 482, L9

\bibitem[{{Pascoe} {et~al.}(2018){Pascoe}, {Anfinogentov}, {Goddard}, \&
  {Nakariakov}}]{2018ApJ...860...31P}
{Pascoe}, D.~J., {Anfinogentov}, S.~A., {Goddard}, C.~R., \& {Nakariakov},
  V.~M. 2018, \apj, 860, 31

\bibitem[{{Pascoe} \& {De Moortel}(2014)}]{2014ApJ...784..101P}
{Pascoe}, D.~J., \& {De Moortel}, I. 2014, \apj, 784, 101

\bibitem[{{Pascoe} {et~al.}(2016{\natexlab{a}}){Pascoe}, {Goddard}, \&
  {Nakariakov}}]{2016A&A...593A..53P}
{Pascoe}, D.~J., {Goddard}, C.~R., \& {Nakariakov}, V.~M. 2016{\natexlab{a}},
  \aap, 593, A53

\bibitem[{{Pascoe} {et~al.}(2016{\natexlab{b}}){Pascoe}, {Goddard},
  {Nistic{\`o}}, {Anfinogentov}, \& {Nakariakov}}]{2016A&A...585L...6P}
{Pascoe}, D.~J., {Goddard}, C.~R., {Nistic{\`o}}, G., {Anfinogentov}, S., \&
  {Nakariakov}, V.~M. 2016{\natexlab{b}}, \aap, 585, L6

\bibitem[{{Pascoe} {et~al.}(2016{\natexlab{c}}){Pascoe}, {Goddard},
  {Nistic{\`o}}, {Anfinogentov}, \& {Nakariakov}}]{2016A&A...589A.136P}
---. 2016{\natexlab{c}}, \aap, 589, A136

\bibitem[{{Pascoe} {et~al.}(2012){Pascoe}, {Hood}, {de Moortel}, \&
  {Wright}}]{2012A&A...539A..37P}
{Pascoe}, D.~J., {Hood}, A.~W., {de Moortel}, I., \& {Wright}, A.~N. 2012,
  \aap, 539, A37

\bibitem[{{Pascoe} {et~al.}(2013){Pascoe}, {Hood}, {De Moortel}, \&
  {Wright}}]{2013A&A...551A..40P}
{Pascoe}, D.~J., {Hood}, A.~W., {De Moortel}, I., \& {Wright}, A.~N. 2013,
  \aap, 551, A40

\bibitem[{{Pascoe} {et~al.}(2019){Pascoe}, {Hood}, \& {Van
  Doorsselaere}}]{2019FrASS...6...22P}
{Pascoe}, D.~J., {Hood}, A.~W., \& {Van Doorsselaere}, T. 2019, Frontiers in
  Astronomy and Space Sciences, 6, 22

\bibitem[{{Reale}(2014)}]{2014LRSP...11....4R}
{Reale}, F. 2014, Living Reviews in Solar Physics, 11, 4

\bibitem[{{Roberts}(2000)}]{2000SoPh..193..139R}
{Roberts}, B. 2000, \solphys, 193, 139

\bibitem[{{Ruderman}(2003)}]{2003A&A...409..287R}
{Ruderman}, M.~S. 2003, \aap, 409, 287

\bibitem[{{Ruderman}(2009)}]{2009A&A...506..885R}
---. 2009, \aap, 506, 885

\bibitem[{{Ruderman}(2015)}]{2015SoPh..290..423R}
---. 2015, \solphys, 290, 423

\bibitem[{{Ruderman} \& {Roberts}(2002)}]{2002ApJ...577..475R}
{Ruderman}, M.~S., \& {Roberts}, B. 2002, \apj, 577, 475

\bibitem[{{Ruderman} \& {Roberts}(2006)}]{2006JPlPh..72..285R}
---. 2006, Journal of Plasma Physics, 72, 285

\bibitem[{{Ruderman} \& {Terradas}(2013)}]{2013A&A...555A..27R}
{Ruderman}, M.~S., \& {Terradas}, J. 2013, \aap, 555, A27

\bibitem[{{Schrijver} {et~al.}(2010){Schrijver}, {DeRosa}, \&
  {Title}}]{2010ApJ...719.1083S}
{Schrijver}, C.~J., {DeRosa}, M.~L., \& {Title}, A.~M. 2010, \apj, 719, 1083

\bibitem[{{Selwa} \& {Ofman}(2010)}]{2010ApJ...714..170S}
{Selwa}, M., \& {Ofman}, L. 2010, \apj, 714, 170

\bibitem[{{Selwa} {et~al.}(2011{\natexlab{a}}){Selwa}, {Ofman}, \&
  {Solanki}}]{2011ApJ...726...42S}
{Selwa}, M., {Ofman}, L., \& {Solanki}, S.~K. 2011{\natexlab{a}}, \apj, 726, 42

\bibitem[{{Selwa} {et~al.}(2011{\natexlab{b}}){Selwa}, {Solanki}, \&
  {Ofman}}]{2011ApJ...728...87S}
{Selwa}, M., {Solanki}, S.~K., \& {Ofman}, L. 2011{\natexlab{b}}, \apj, 728, 87

\bibitem[{{Soler} {et~al.}(2013){Soler}, {Goossens}, {Terradas}, \&
  {Oliver}}]{2013ApJ...777..158S}
{Soler}, R., {Goossens}, M., {Terradas}, J., \& {Oliver}, R. 2013, \apj, 777,
  158

\bibitem[{{Soler} {et~al.}(2014){Soler}, {Goossens}, {Terradas}, \&
  {Oliver}}]{2014ApJ...781..111S}
---. 2014, \apj, 781, 111

\bibitem[{{Soler} \& {Terradas}(2015)}]{2015ApJ...803...43S}
{Soler}, R., \& {Terradas}, J. 2015, \apj, 803, 43

\bibitem[{{Spruit}(1982)}]{1982SoPh...75....3S}
{Spruit}, H.~C. 1982, \solphys, 75, 3

\bibitem[{{Terradas} {et~al.}(2006{\natexlab{a}}){Terradas}, {Oliver}, \&
  {Ballester}}]{2006ApJ...650L..91T}
{Terradas}, J., {Oliver}, R., \& {Ballester}, J.~L. 2006{\natexlab{a}}, \apjl,
  650, L91

\bibitem[{{Terradas} {et~al.}(2006{\natexlab{b}}){Terradas}, {Oliver}, \&
  {Ballester}}]{2006ApJ...642..533T}
---. 2006{\natexlab{b}}, \apj, 642, 533

\bibitem[{{Tian} {et~al.}(2012){Tian}, {McIntosh}, {Wang}, {Ofman}, {De
  Pontieu}, {Innes}, \& {Peter}}]{2012ApJ...759..144T}
{Tian}, H., {McIntosh}, S.~W., {Wang}, T., {et~al.} 2012, \apj, 759, 144

\bibitem[{{Tomczyk} \& {McIntosh}(2009)}]{2009ApJ...697.1384T}
{Tomczyk}, S., \& {McIntosh}, S.~W. 2009, \apj, 697, 1384

\bibitem[{{Tomczyk} {et~al.}(2007){Tomczyk}, {McIntosh}, {Keil}, {Judge},
  {Schad}, {Seeley}, \& {Edmondson}}]{2007Sci...317.1192T}
{Tomczyk}, S., {McIntosh}, S.~W., {Keil}, S.~L., {et~al.} 2007, Science, 317,
  1192

\bibitem[{{Van Doorsselaere} {et~al.}(2004){Van Doorsselaere}, {Andries},
  {Poedts}, \& {Goossens}}]{2004ApJ...606.1223V}
{Van Doorsselaere}, T., {Andries}, J., {Poedts}, S., \& {Goossens}, M. 2004,
  \apj, 606, 1223

\bibitem[{{Van Doorsselaere} {et~al.}(2008){Van Doorsselaere}, {Nakariakov},
  {Young}, \& {Verwichte}}]{2008A&A...487L..17V}
{Van Doorsselaere}, T., {Nakariakov}, V.~M., {Young}, P.~R., \& {Verwichte}, E.
  2008, \aap, 487, L17

\bibitem[{{Van Doorsselaere} {et~al.}(2009){Van Doorsselaere}, {Verwichte}, \&
  {Terradas}}]{2009SSRv..149..299V}
{Van Doorsselaere}, T., {Verwichte}, E., \& {Terradas}, J. 2009, \ssr, 149, 299

\bibitem[{{Verth} \& {Erd{\'e}lyi}(2008)}]{2008A&A...486.1015V}
{Verth}, G., \& {Erd{\'e}lyi}, R. 2008, \aap, 486, 1015

\bibitem[{{Verwichte} {et~al.}(2009){Verwichte}, {Aschwanden}, {Van
  Doorsselaere}, {Foullon}, \& {Nakariakov}}]{2009ApJ...698..397V}
{Verwichte}, E., {Aschwanden}, M.~J., {Van Doorsselaere}, T., {Foullon}, C., \&
  {Nakariakov}, V.~M. 2009, \apj, 698, 397

\bibitem[{{Verwichte} {et~al.}(2004){Verwichte}, {Nakariakov}, {Ofman}, \&
  {Deluca}}]{2004SoPh..223...77V}
{Verwichte}, E., {Nakariakov}, V.~M., {Ofman}, L., \& {Deluca}, E.~E. 2004,
  \solphys, 223, 77

\bibitem[{{Verwichte} {et~al.}(2013){Verwichte}, {Van Doorsselaere}, {White},
  \& {Antolin}}]{2013A&A...552A.138V}
{Verwichte}, E., {Van Doorsselaere}, T., {White}, R.~S., \& {Antolin}, P. 2013,
  \aap, 552, A138

\bibitem[{{Vigeesh} {et~al.}(2009){Vigeesh}, {Hasan}, \&
  {Steiner}}]{2009A&A...508..951V}
{Vigeesh}, G., {Hasan}, S.~S., \& {Steiner}, O. 2009, \aap, 508, 951

\bibitem[{{Wang} {et~al.}(2018){Wang}, {Deng}, {Li}, {Feng}, {Bai}, {Deng},
  {Yang}, {Xue}, \& {Wang}}]{2018ApJ...856L..16W}
{Wang}, F., {Deng}, H., {Li}, B., {et~al.} 2018, \apjl, 856, L16

\bibitem[{{Wang} \& {Sakurai}(1998)}]{1998PASJ...50..111W}
{Wang}, H., \& {Sakurai}, T. 1998, \pasj, 50, 111

\bibitem[{{Wang} {et~al.}(2012){Wang}, {Ofman}, {Davila}, \&
  {Su}}]{2012ApJ...751L..27W}
{Wang}, T., {Ofman}, L., {Davila}, J.~M., \& {Su}, Y. 2012, \apjl, 751, L27

\bibitem[{{Wang}(2016)}]{2016GMS...216..395W}
{Wang}, T.~J. 2016, Washington DC American Geophysical Union Geophysical
  Monograph Series, 216, 395

\bibitem[{{Wentzel}(1979)}]{1979ApJ...227..319W}
{Wentzel}, D.~G. 1979, \apj, 227, 319

\bibitem[{{White} \& {Verwichte}(2012)}]{2012A&A...537A..49W}
{White}, R.~S., \& {Verwichte}, E. 2012, \aap, 537, A49

\bibitem[{{Wright} \& {Thompson}(1994)}]{1994PhPl....1..691W}
{Wright}, A.~N., \& {Thompson}, M.~J. 1994, Physics of Plasmas, 1, 691

\bibitem[{{Zimovets} \& {Nakariakov}(2015)}]{2015A&A...577A...4Z}
{Zimovets}, I.~V., \& {Nakariakov}, V.~M. 2015, \aap, 577, A4

\end{thebibliography}
		
\end{document}